\documentclass[fleqn,usenatbib]{mnras}

\usepackage[T1]{fontenc}
\usepackage{ae,aecompl}

\usepackage{graphicx}	% Including figure files
\usepackage{amsmath}	% Advanced maths commands
\usepackage{amssymb}	% Extra maths symbols

\usepackage{graphicx,color}	% Including figure files

\newcommand{\kms}{\textrm{km s}^{-1}}
\newcommand{\hi}{\text{H\,\sc{i}}}

\def\bsq#1{%both single quotes
\lq{#1}\rq}

\title[Asymmetries, max. 45 characters]{Systematically Asymmetric: A comparison of \hi\ profile asymmetries in real and simulated galaxies.}

\author[N. Deg et al.]{
N. Deg,$^{1,2}$\thanks{E-mail: nathan.deg@ast.uct.ac.za}
S.-L. Blyth$^{2}$, N. Hank$^{2}$, S. Kruger$^{2}$, 
C. Carignan$^{2,3}$
\\
$^{1}$Department of Physics, Engineering Physics, and Astronomy, Queen's University, Kingston, ON, K7L 3N6, Canada\\
$^{2}$Department of Astronomy, University of Cape Town, Private Bag X3, Rondebosch 7701, South Africa\\
$^{3}$Observatoire d'Astrophysique de l'Universite de Ouagadougou (ODAUO), BP 7021, Ouagadougou 03, Burkina Faso
}

\date{Accepted XXX. Received YYY; in original form ZZZ}

\pubyear{2017}

\begin{document}
\label{firstpage}
\pagerange{\pageref{firstpage}--\pageref{lastpage}}
\maketitle

\begin{abstract}

We examine different measures of asymmetry for galaxy HI velocity profiles. We introduce 
the channel-by-channel asymmetry  and the velocity-of-equality statistics to quantify 
profile asymmetries. Using a sample of simulated galaxies, we examine
 how these and the standard lopsidedness morphometric statistic 
 depend on a variety of observational effects including the viewing angle 
 and inclination. We find that our newly introduced 
 channel-by-channel asymmetry is less sensitive to the effects of 
 viewing angle and inclination than other morphometrics. 
 Applying our statistics to the WHISP HI galaxy sample, we also find that the 
 channel-by-channel asymmetry, is a better indicator of visually-classified 
 asymmetric profiles. In addition, we find that the 
 lopsidedness-velocity of equality space can be used to identify profiles with deep
  central dips without visual inspection. 
\end{abstract}


%%%%%%%%%%%%%%%%%%%%%%%%%%%%%%%%%%%%%%%%%%%%%%%%%%

%%%%%%%%%%%%%%%%% BODY OF PAPER %%%%%%%%%%%%%%%%%%

\section{Introduction}

Galaxies often appear axisymmetric on large scales,
but a closer examination usually reveals a variety
of asymmetric features.  These
include tidal tails, various stripping effects, 
off-centered
spiral arms, and lopsidedness.  At the smallest 
scales, even star formation can appear asymmetric.
These asymmetries are often more pronounced in the
gaseous component of galaxies, as it extends further than the stellar component.
A key driver of morphological and kinematic 
disturbances that results in asymmetries are 
galaxy interactions with their environments and 
neighbours \citep{Keres2002, FernandezLorenzo2013, Mundy2017}. From both observations and modelling of the stellar and gas components of galaxies, it has been seen that galaxy-galaxy interactions and mergers \citep{Reichard2008}, galaxy-environment interactions \citep{Angiras2006, vanEymeren2011} , ram-pressure stripping, \citep{Gunn1972} and gas accretion from the cosmic web \citep{Bournard2005} can 
all drive morphological and kinematic asymmetries. The relative importance of these processes in different environments and at different epochs is not yet understood and is still an open question in the field of galaxy evolution. Studying the stellar and gas asymmetries of galaxies in these different environments will help to provide information on the processes driving their evolution.

Various photometric techniques have been developed over the years to quantify two-dimensional (2D) morphological asymmetries based on galaxy stellar light distributions. The most well-known are the concentration-asymmetry-smoothness (CAS) parameters \citep{Trujillo2001, Conselice2003}, the $M_{20}$ parameter, and the $G_{ini}$ coefficient \citep{Lotz2004}. 
These statistics were initially designed to develop a 
non-parametric set of classifications for 2D images. 
\citet{Conselice2000} found that a combination of 
color and asymmetry could be used to distinguish between
spirals, ellipticals, and edge-on galaxies.  
\citet{Conselice2003} followed up on this work and found that
these morphological statistics correlate with important
physical features.
The concentration correlates with B/T ratios, 
smoothness tends to measure H$\alpha$ 
equivalent widths, and asymmetry can
identify galaxies undergoing major mergers.

The idea of using morphological statistics as a 
classification method has been used in a variety of different
areas.  For instance, asymmetry has been used to find 
merging galaxies in 
Integral field unit (IFU) spectroscopy surveys of 
objects at intermediate to high 
redshift \citep{Shapiro2008, Bloom2017, Bloom2018, Wisnioski2019}.

The cold gas component, in particular the neutral hydrogen, \hi, in galaxies is also important to take into account when investigating the effects of galaxy interactions. Typically, in disk galaxies, the \hi\ is a significant fraction of the total mass of the galaxy and also has a more extended radial profile than the stellar component. Due to its nature and \textcolor{black}{more extended radial} distribution, the gas is therefore more susceptible to disruption \textcolor{black}{on earlier timescales} in galaxy-galaxy interactions than the stellar component of a galaxy. \textcolor{black}{Asymmetries in the HI distribution and kinematics may also be caused by accretion of gas onto galaxies from the cosmic web and/or galactic fountain processes (for a review of HI accretion in galaxies see \cite{Sancisi2008}).}

\cite{Kornreich2000} investigated the asymmetries of 9 face-on spiral galaxies based on \hi\ imaging data using several different techniques. More recently, various groups \citep{Holwerda2011, Lelli2014, Giese2016} have applied quantitative methods similar to those employed in optical studies to measure the 2-D asymmetries of galaxy \hi\ images. However, sample sizes of existing \hi\ galaxy imaging surveys are of the order of at most a few hundred galaxies and are limited to the local Universe. For instance, the 
Westerbork HI survey of Irregular and SPiral galaxies (WHISP)
has a catalogue of 375 galaxies \citep{vanderHulst2001}.
Compared to optical surveys, the sample sizes of \hi\ imaging surveys are much smaller and the redshift coverage has been constrained to the local Universe due to the extremely long observing times required for radio observations on interferometers (although 
upcoming surveys using MeerKAT and ASKAP will
soon provide large numbers
of resolved HI images). 
Single-dish surveys on the other hand, for example, HIPASS \citep{Meyer2004} and ALFALFA \citep{Haynes2018}, have observed the global \hi\ profiles of thousands of galaxies. 

Typically \hi\ profiles are characterized by three numbers;
the integrated \hi\ flux, the systemic velocity, and the 
profile width.  Asymmetry gives at least one new statistic
that can be used to characterize galaxies.  
Galaxy \hi\ asymmetries have to date typically been measured in a range of different environments in the local universe using the galaxies' \hi\ profiles. In 1974, \cite{Peterson1974} measured the asymmetry of the \hi\ profiles of a sample of galaxies from the \textit{Atlas of Peculiar Galaxies} by \cite{Arp1966} and found that the \hi\ profile asymmetries correlated with the optical morphological asymmetries they observed. Since then, many studies have measured \hi\ profile asymmetries based on single-dish data; \cite{Tifft1988} characterised \hi\ profiles observed with the 91 m NRAO telescope at Greenbank and reported asymmetries calculated as a ratio of velocity differences between the edges of a profile and the velocity value resulting in equal area on each side under the profile. By analysing a large sample of $\sim1700$ mostly field galaxy \hi\ profiles, by eye, \cite{Richter1994} concluded that approximately half of galaxies have asymmetric \hi\ profiles. Using the Green Bank 43m telescope \cite{Haynes1998} observed the \hi\ profiles of 78 isolated spiral galaxies and used two quantitative methods to measure the profile lopsidedness. Classifying profiles as lopsided if the difference in area under the profile on each side of the median systemic velocity was greater than 5 percent resulted in a similar asymmetry rate of $\sim50$ percent for their sample. \cite{Matthews1998} found a $\sim$77 percent \hi\ profile asymmetry rate in their sample of 30 late-type spirals and also found a generally higher rate of asymmetry in the \hi\ profiles compared to the optical images for their sample. 

More recently, \cite{Espada2011} set out to measure the \hi\ profile lopsidedness distribution of extremely isolated galaxies in the AMIGA (Analysis of the Interstellar Medium of Isolated GAlaxies) sample\footnote{http://amiga.iaa.es} in order to estimate a baseline intrinsic asymmetry rate against which other samples in different environments could be compared. They used the same quantitative technique of comparing the integrated flux ratios on each side of the median systemic velocity. Fitting a half-Gaussian distribution to their data resulted in a standard deviation of $\sigma=0.13$ and galaxies were classified as asymmetric if their lopsidedness values deviated by more than 2$\sigma$, i.e., had lopsidedness values $A>1.26$. Using this asymmetry criterion, only 9 percent of the AMIGA \hi-refined sample of 166 galaxies were found to be asymmetric and applying the same criterion to previous studies, they found a similar rate for the isolated \cite{Haynes1998} sample but a larger rate for the \cite{Matthews1998} field sample (see Table 3 in their paper \citep{Espada2011}). 

To probe the impact of mergers on galaxy \hi\ profile asymmetries, \cite{Bok2019} investigated the rate of lopsidedness in ALFALFA \citep{Haynes2018} \hi\ profiles of 348 galaxies in close pairs (an optical 
neighbour within 100 kpc and 500 $\kms$) identified in the Sloan Digital Sky Survey DR7 \citep{Abazajian2009} using the same flux ratio method. Based on the \cite{Espada2011} asymmetry criterion, they found a significantly higher asymmetry rate (27 percent) in their close pair sample compared to the isolated and field samples of \cite{Espada2011}, \cite{Haynes1998}, and \cite{Matthews1998} and their own field sample (18 percent), indicating statistically that merger activity is a driver of \hi\ profile lopsidedness.

\citet{Watts2020} recently examined a sample of 562 galaxies from the 
extended GALEX Arecibo SDSS Survey (xGASS) \citep{Catinella2010, Catinella2018}.
They also found that environment was a key driver of asymmetries, with satellite galaxies
having a higher rate of asymmetry than central galaxies, and that isolated centrals 
had slightly higher symmetry than group centrals.  Interestingly, they 
also found that asymmetric galaxies in their sample typically had a lower gas fraction
than symmetric galaxies.

In the near future, thousands of galaxy \hi\ images are expected to be available
 from SKA precursor shallow \hi\ surveys such as WALLABY \citep{WALLABYDINGO} on ASKAP  
and MIGHTEE \citep{MIGHTEE} on MeerKAT. However, the deeper \hi\ surveys such 
as DINGO  \citep{WALLABYDINGO} on ASKAP and LADUMA \citep{LADUMA} 
on MeerKAT that are aiming to observe \hi\ in galaxies far beyond the local universe, 
will not spatially resolve the \hi\ disks of higher redshift galaxies.  An interesting
aspect of asymmetry measurements is, due to the channel
resolution being roughly equal at low and intermediate redshifts (z$\sim$0 and z$\sim$ 0.5 respectively),
a direct comparison of profile asymmetries at various
redshifts is possible.   Moreover, just as
the 2D morphological statistics can be used to classify 
galaxies \citep{Conselice2003}, it may be possible to 
classify profiles with a full suite of different 
asymmetry measurements. 
\textcolor{black}{While most of the methods (including those presented in this paper) used to quantify 2D and 1D asymmetries in galaxies are non-parametric, it is also possible to extract asymmetry measures by fitting HI global profiles with physically motivated models (e.g. \cite{Stewart2014}) or parameterised functions (e.g. the ‘Busy function’ \cite{Westmeier2014}). Since parametric methods have not yet been broadly applied in the literature, we focus rather on comparisons to non-parametric methods here.}

In this paper, with both the upcoming 
\hi\ datasets from the SKA pathfinders 
and the idea of characterizing velocity profiles with 
multiple statistics in mind, 
we introduce two new asymmetry statistics.
One of these is a 1D analogue of the 2D asymmetry statistic, 
while the other  is similar to the velocity method 
given by \citet{Haynes1998}, but is derived from 
the profile lopsidedness.  We examine how these
depend on a variety of observational effects using 
mock profiles and numerical simulations.  We also 
calculate these statistics for the WHISP sample and 
compare them to a visual classification of the profiles.

In Section \ref{Sec:AsymmetryMeasures}
we present statistics for quantifying velocity 
profiles.  Section \ref{Sec:SimpleProfiles} explores 
the dependence of these statistics on the \bsq{folding}
velocity and the S/N using mock test
profiles.  
Section \ref{Sec:Simulations} presents two simulations
of interacting galaxy pairs.  These are used to explore how 
the various statistics depend on the observed viewing 
angle and inclination.  Next, in Sec. \ref{Sec:WHISP},
we apply these statistics to a
sample of galaxies from the 
WHISP survey.  
Finally, Sec. \ref{Sec:Conclusions} presents
our discussion and conclusions about these statistics.

\section{Measures of Asymmetry}\label{Sec:AsymmetryMeasures}

Compared to the variety of two-dimensional methods, there are considerably few statistics for analyzing 
1-D velocity profiles.  The most common is the 
lopsidedness statistic which compares the 
flux in the approaching and receding portions of a
particular velocity profile (e.g. \cite{Haynes1998}).  

It is important to make a few notes on terminology here.
Typically lopsidedness 
and asymmetry are used interchangeably when discussing velocity
profiles.  In this work we will introduce other 
asymmetry statistics so we will exclusively call
the usual statistic lopsidedness.  One of the other 
statistics that we will introduce involves a 
channel-by-channel comparison of fluxes that is 
similar in construction to the 2-D asymmetry 
statistic.  For the sake of comparison we will 
refer to that statistic as the channel-by-channel
asymmetry or as the asymmetry statistic.

\textcolor{black}{
In addition, this work uses a variety of different
characteristic velocities that are introduced throughout
the paper.  To that end, \textcolor{black}{Table \ref{Tab:Velocities}} summarizes the 
different velocity measures along with where they 
are defined.
}
\begin{table*}
    \centering
    \begin{tabular}{c|c|l|}
        \hline 
        \textcolor{black}{Symbol} & \textcolor{black}{Location} & \textcolor{black}{Definition} \\
        \hline
         \textcolor{black}{$v_{low}$} & \textcolor{black}{Sec. \ref{subsec:Lopsidedness}} & \textcolor{black}{The
         smallest velocity in the profile given the chosen width.} \\
        \textcolor{black}{$v_{high}$} & \textcolor{black}{Sec. \ref{subsec:Lopsidedness}} & \textcolor{black}{The
         highest velocity in the profile given the chosen width.} \\
        \textcolor{black}{$v_{fold}$} & \textcolor{black}{Sec. \ref{subsec:Lopsidedness}} & \textcolor{black}{The
         velocity used to separate the lower portion of}\\
         & & \textcolor{black}{the profile from the higher when calculating various asymmetry statistics.} \\
        \textcolor{black}{$v_{sys}$} & \textcolor{black}{Sec. \ref{subsec:VEqual}} & \textcolor{black}{The
         systemic velocity defined as the midway point between $v_{low}$ and $v_{high}$.} \\
        \textcolor{black}{$v_{equal}$} & \textcolor{black}{Sec. \ref{subsec:VEqual}} & \textcolor{black}{The
         value of $v_{fold}$ that gives zero lopsidedness.} \\
        \textcolor{black}{$v_{l,i}$} & \textcolor{black}{Sec. \ref{SubSec:AsymmetryDef}} & \textcolor{black}{The
         low velocity channel for a velocity pair.} \\
        \textcolor{black}{$v_{h,i}$} & \textcolor{black}{Sec. \ref{SubSec:AsymmetryDef}} & \textcolor{black}{The
         high velocity channel for a velocity pair.} \\
        \textcolor{black}{$v_{sym}$} & \textcolor{black}{Sec. \ref{subsec:VfoldResponse}} & \textcolor{black}{The
         folding velocity that minimizes $\mathcal{A}$.} \\
    \end{tabular}
    \caption{\textcolor{black}{A list of the 
    different velocities used in this work.}}
    \label{Tab:Velocities}
\end{table*}

\subsection{Lopsidedness}\label{subsec:Lopsidedness}

\begin{figure}
\centering
    \includegraphics[width=70mm]{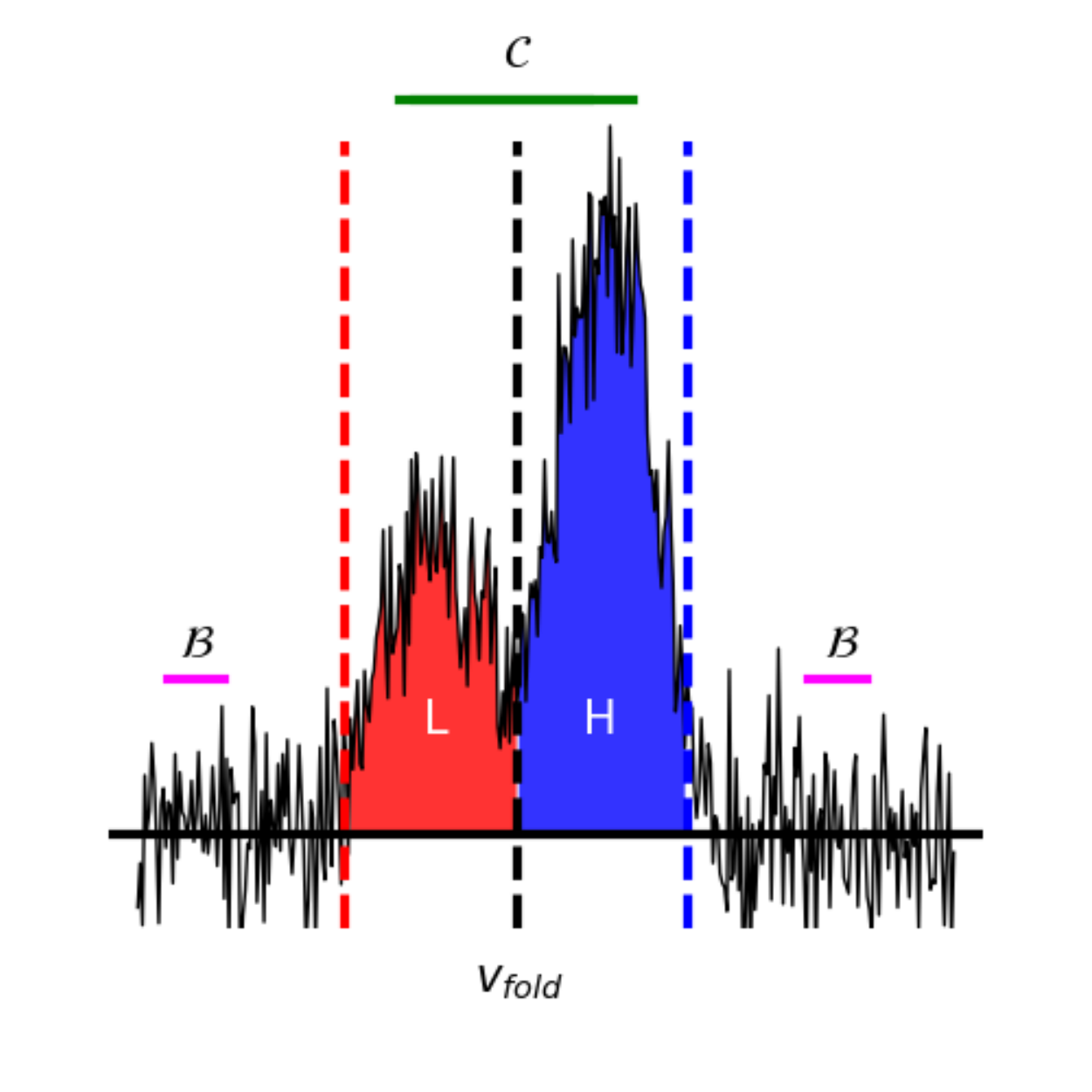} 
\caption{A graphical explanation of the lopsidedness and 
asymmetry statistics. The lopsidedness compares the red
and blue areas.  The lopsidedness can be set to either the maximal
ratio of $L$ and $H$, $A_{L,r}$ or to the normalized difference
between the two, $A_{L}$.
The channel-by-channel asymmetry statistic, $\mathcal{A}$ compares the difference
between pairs of matched channels, which are then 
summed to $\mathcal{C}$.  A similar number of channels beyond the
range of the profile are also compared to get $\mathcal{B}$, which
is then subtracted from $\mathcal{C}$ to get $\mathcal{A}$.
The dashed red and blue vertical lines indicate the edge of the 
velocity profile while the dashed black line indicates
the folding velocity used for the calculation of $A_{L}$ and 
$\mathcal{A}$.}
  \label{Fig:TestProfile}
\end{figure}

The lopsidedness statistic compares the integrated 
flux on either side of some \bsq{folding} velocity, $v_{fold}$.
It is typically set equal to the systemic\textcolor{black}{/median} velocity of the
galaxy, $v_{sys}$. In the 
literature the lopsidedness is most often calculated 
(e.g. \citet{Haynes1998,Espada2011,Bok2019}) as
\begin{equation}\label{Eq:Lopsidedness}
    A_{L,r}=\textrm{max}\left(\frac{H}{L},\frac{L}{H} \right)~,
\end{equation}
where $A_{L,r}$ is the lopsidedness ratio
and $L$ and $H$ are the integrated 
flux on the low/high side of 
the profile respectively.  Explicitly, this is
\begin{equation}
    L=\int_{v_{low}}^{v_{fold}}f(v)dv~,
\end{equation}
and
\begin{equation}
    H=\int_{v_{fold}}^{v_{high}}f(v)dv~,
\end{equation}
where $f(v)$ is the flux density as a function of velocity
and $v_{fold}$ is the velocity separating
$L$ from $H$.  Figure
\ref{Fig:TestProfile} shows a sample velocity profile to 
illustrate the lopsidedness statistic.

The limits on Eq. \ref{Eq:Lopsidedness} are formally 
one and infinity, where 
one is perfectly symmetric and infinity is completely 
lopsided.  In practice, a lopsidedness of two or more is 
quite extreme.  However, the 2-D analogues of lopsidedness
have limits between zero and one.  As such, we prefer the definition of lopsidedness developed by \cite{Peterson1974}.
In their definition, 
\begin{equation}\label{Eq:NewLopsidedness}
    A_{L}=\frac{|L-H |}{L+H}~.
\end{equation}
In this formulation, $0 \le A_{L} \le 1$, where $A_{L}=0$ 
is perfectly 
symmetric.  It is straightforward to convert between 
Eq. \ref{Eq:Lopsidedness} and \ref{Eq:NewLopsidedness} using
\begin{equation}
    A_{L}=\frac{A_{L,r}-1}{A_{L,r}+1}~.
\end{equation}
In terms of 
true dynamical range, $A_{L}=0.33$ corresponds to a flux 
ratio of 2, which, as mentioned earlier, is very asymmetric.

\subsection{Velocity of Equality}\label{subsec:VEqual}

Typically the lopsidedness statistic is calculated at 
the systemic velocity of the profile.  To be clear, the
systemic velocity we use is 
$v_{sys}=(v_{high}-v_{low})/2$ where $v_{high}$ and $v_{low}$
are the edge velocities, which in turn are defined by the peak to edge 
flux ratio (see Sec. \ref{Sec:Implementation}). For single peaked profiles
$F(v_{low~peak})$=$F(v_{high~peak})$, but for 
double horned profiles, these fluxes can be different.

That being said, in every profile
there exists a
folding velocity such that $A_{L}(v_{equal})=0$,
where $v_{equal}$ is the
\bsq{velocity of equality}.  This velocity can 
be used to define a new statistic, $\Delta V$.  This is
the difference between the systemic velocity and the 
velocity of equality normalized by the width of the 
profile.  Explicitly, this is
\begin{equation}\label{Eq:DeltaV}
    \Delta V= \frac{2|v_{sys}-v_{equal}|}{w}~,
\end{equation}
where $w$ is the width of the velocity profile (defined by the profile edges).  
The factor of 2 is introduced so that the formal limits on this statistic are $0 \le \Delta V \le 1$,
but generally the magnitude will be significantly smaller. 

Our $\Delta V$ statistic is very similar to how 
\citet{Haynes1998} used differences between the 
flux-weighted mean velocity, $\bar{v}$, 
and the systemic velocity to characterize
profile asymmetries.  The key difference between our two 
methods is that $v_{equal} \ne \bar{v}$. The 
flux-weighted mean velocity
does not always coincide with the 
velocity where the lopsidedness is zero.  While the 
two statistics are similar, we use  $\Delta V$ in this work 
as it is explicitly connected to $A_{L}$ through $v_{equal}$.

It is worth noting
that the asymmetry statistic used by \citet{Tifft1988} also 
uses $v_{equal}$, but the formulation is quite different than
our $\Delta V$.  They examine the ratio of $(v_{equal}-v_{low})$
and $(v_{high}-v_{equal})$ and $v_{sys}$ is only included to give the 
sign and determine which factor will be the numerator or denominator.  It
is closer in construction to Eq. \ref{Eq:Lopsidedness} than Eq. \ref{Eq:DeltaV}.

\subsection{Channel-by-Channel Asymmetry}\label{SubSec:AsymmetryDef}

The lopsidedness statistic is a relatively global 
statistic.  It is an integral quantity that compares the total flux on 
either side of some central velocity.  However, the 
2-D asymmetry statistic is much more local \citep{Conselice2000}.  
The 2-D asymmetry is
\begin{equation}
    \mathcal{A}_{2D}= \frac{\sum_{i}|f_{i}-f_{180,i }|}
    {2\sum_{i}f_{i}}~,
\end{equation}
where $f_{i}$ is the flux of a pixel and
$f_{180,i}$ is the flux of the pixel located at 
$180^{\circ}$ rotation from the $i$'th pixel.  The 
factor of 2 in the denominator is present to 
deal with double-counting.  

The difference between the integrated lopsidedness and the more local 2-D asymmetry
has, in part, motivated the development of a new channel-by-channel
1-D asymmetry statistic, $\mathcal{A}$.  
Related to this motivation is the fact that lopsidedness integrates
out small, dynamically local variations in the velocity
profile.  These variations may contain subtle information
about the galaxy that lopsidedness will not detect.

The channel-by-channel asymmetry (or 1-D 
asymmetry due to it's similarity to the 2-D asymmetry)
is the normalized sum of 
flux differences across $v_{fold}$.  Because this 
quantity is local, it may be strongly affected by
background noise.  As such, we calculate a
background term and subtract it from the 
summation in the signal region to obtain 
the measured asymmetry. 
Explicitly
\begin{equation} \label{Eq:Asymmetry}
    \mathcal{A}_{meas}=\mathcal{C}-\mathcal{B}
\end{equation}
where $\mathcal{C}$ is the degree of asymmetry 
in the profile, while $\mathcal{B}$ is the 
amount of asymmetry due to noise.  The 
signal term is
\begin{equation}\label{Eq:1D_Asymmetry_NoBack}
\mathcal{C}=\frac{\sum_{i}|f(v_{l,i})\delta 
v-f(v_{h,i})\delta v|}
{\sum_{i}\left((f(v_{l,i})\delta v
+f(v_{h,i})\delta v\right)}~,
\end{equation}
where $v_{l,i}$ and $v_{h,i}$ are the $i$'th velocity 
pair and $\delta v$ is the channel width.  
They are given simply by $v_{(l,h),i}=v_{sys} \pm (i \times \delta v)$.
The background term is similar to the profile
term except it is still normalized by 
the total flux density of the velocity profile.  That is
\begin{equation}
    \mathcal{B}=\frac{\sum_{i}|f(v_{b,l,i})\delta v-f(v_{b,h,i})\delta v|}
{\sum_{i}\left(f(v_{l,i})\delta v+f(v_{h,i})\delta v\right)}~,
\end{equation}
where $f(v_{b,i})$ is the flux density in some 
channel outside 
the velocity profile. 
Figure \ref{Fig:TestProfile} shows the location of 
$\mathcal{C}$ and $\mathcal{B}$ relative to a sample velocity 
profile. 

By construction, the limits on $\mathcal{C}$ and $\mathcal{B}$
are between zero and one.  Since $\mathcal{A}$ is 
the difference of the signal and background terms, it is 
possible for it to be negative.  This implies 
that the profile is noise dominated and a 
channel-by-channel asymmetry measurement 
cannot be made reliably.

The 1-D asymmetry and lopsidedness statistics are similar in
both construction and range, but they measure different
things. \textcolor{black}{While both statistics are 
calculated on the global HI profile, the channel-by-channel
asymmetry is sensitive to \bsq{local} perturbations (where 
local refers to specific velocity channels).  Lopsidedness
is more sensitive to larger scale perturbations to the 
velocity profile.}

\textcolor{black}{As a thought experiment, it is possible to 
imagine observing a \bsq{real} galaxy 
using an instrument with 
infinite velocity resolution and no noise.  The galaxy will have a symmetric 
double horned profile with zero lopsidedness.  
For there to be zero channel-by-channel asymmetry the 
infinite resolution of the instrument requires that
the galaxy consist of symmetrically arranged gas clouds
such that each approaching cloud is paired with a receding
cloud with equal and opposite velocity relative to 
$v_{sys}$. 
In a \bsq{real} galaxy, gas
clouds are drawn from a distribution function and
such pairs are
unlikely.  Instead this galaxy \bsq{observation}
would tend towards
having $\mathcal{A}=1$ as each channel pair would 
contain the flux of a single gas cloud (again, this
is due to the infinite resolution and discretized gas clouds).
This is a contrived example, but it illustrates
that the lopsidedness and 
channel-by-channel asymmetry statistics measure different
things.}

\subsection{Implementation}\label{Sec:Implementation}

We have implemented our profile analysis using custom 
\textsc{python} code.  The code itself uses the standard \textsc{numpy}
package for most array calculations.  In some 
later steps, we utilize the \textsc{lmfit}
\citep{LMFIT} 
\textcolor{black}{\textsc{python} package to calculate} the \bsq{velocity of symmetry}
(see \textcolor{black}{Tab. \ref{Tab:Velocities} and}
Sec. \ref{Sec:SimpleProfiles} for
details).  

The lopsidedness integrations are 
performed using
a modified trapezoid integral that can account for 
unequal channel spacings.  This is necessary as neither
the edges nor $v_{fold}$ may lie directly on a 
particular channel value.  

The calculation of $v_{equal}$ for 
$\Delta V$ uses a simple bisection rootfinder
on a version of Eq. \ref{Eq:NewLopsidedness}
that does not include an absolute sign.  The
\bsq{signed} lopsidedness has limits of one and 
negative one and passes through zero at 
$v_{equal}$.

The channel-by-channel asymmetry is calculated
by taking pairs of $v_{i}$ in equal
outward steps of $\delta v$ from
$v_{fold}$.  When
$v_{fold}$ does not lie on a channel, we use 
linear interpolation to find the flux values for
each $v_{i}$ in Eq. \ref{Eq:1D_Asymmetry_NoBack}.

More importantly, a technical issue arises
when $v_{fold} \ne v_{sys}$.  $\mathcal{A}$
requires an equal number of channels on either side of 
$v_{fold}$.  In order to include the entire profile, it 
is necessary to consider channels beyond the edge.
In other words
the width considered, $w_{\mathcal{A}}$, is 
\begin{equation}
    w_{\mathcal{A}}=2\textrm{max}
    (v_{fold}-v_{low},v_{high}-v_{fold})~,
\end{equation}
and is centered on $v_{fold}$.

As a note, in the rest of this work, we define the 
profile edges as the velocities where the flux equals
$20\%$ of the peak flux or fluxes for 
single or double peaked profiles respectively.

%\section{Simple Profiles} \label{Sec:SimpleProfiles}
\section{Measurement Systematics} \label{Sec:SimpleProfiles}

In order to test the dependence of the lopsidedness 
and asymmetry statistics on the folding velocity, signal-to-noise, and resolution, 
we have generated a set of mock velocity profiles
from the sums of Gaussian functions.  Since these
profiles are simple and noise free, they 
allow for an isolation of the effect of the 
folding velocity.  The set of profiles consists
of a single-peaked Gaussian, $G$, a symmetric double-peaked
profile, $S$, a profile that is slightly asymmetric, \textit{SA},
a very asymmetric profile, \textit{VA}, and a fifth profile that has
one broad low-flux peak and one thin high-flux peak, \textit{BT}.

\subsection{Dependence on Folding Velocity}\label{subsec:VfoldResponse}
\begin{figure*}
\centering
    \includegraphics[width=150mm]{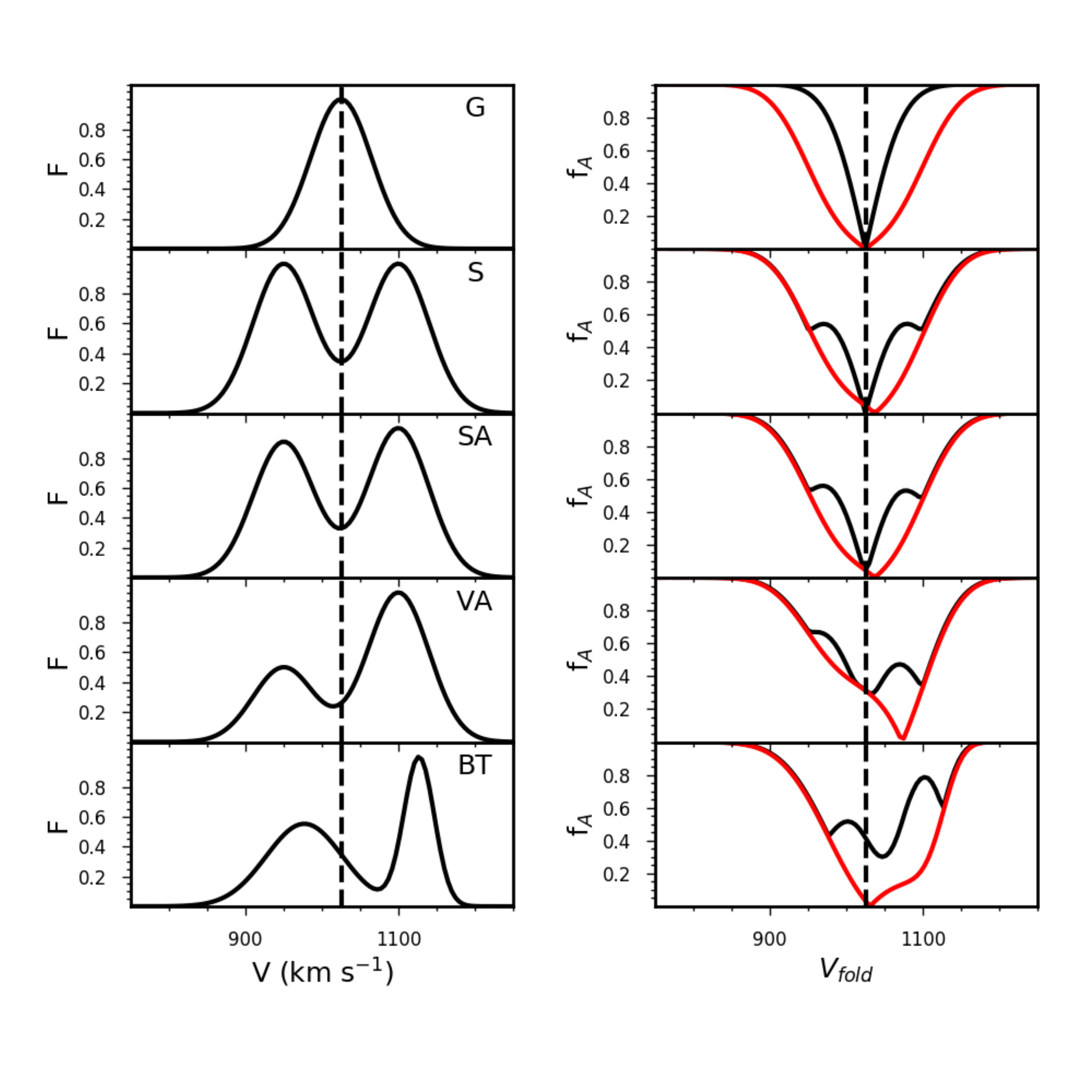} 
\caption{Normalized test velocity profiles (left panels)
and the 
asymmetry (black lines) and lopsidedness (red lines)
statistics as a function 
of $v_{fold}$ for each profile (right panels). The 
five velocity profiles are a Gaussian profile, $G$,
a symmetric double peaked profile, $S$, a
slightly asymmetric double-peaked profile, \textit{SA}, 
a very asymmetric double-peaked profile, \textit{VA}, and
a profile with one broad peak and one thin
peak, \textit{BT}.}
  \label{Fig:VSysResponse}
\end{figure*}

Figure \ref{Fig:VSysResponse} 
shows the lopsidedness and asymmetry statistics
as a function of $v_{fold}$.  As noted in the 
discussion of $v_{equal}$, $A_{L}$ always has a minimum 
of zero.  This does not always occur at 
$v_{sys}$, leading to our proposed $\Delta V$ statistic.

It is abundantly clear in Fig. \ref{Fig:VSysResponse} that 
the channel-by-channel asymmetry statistic (shown in black)
has a different profile than the lopsidedness (shown in red).  In all 
the double-peaked profiles, there is one global minimum, 
and, for the double peaked profiles, two local 
minima that occur at the locations of the peaks. 
These local minima occur due to the
symmetry of the peaks cancelling out pairs near
those values of $v_{fold}$.

Unlike the lopsidedness, the asymmetry statistic has a 
unique minimum for each profile.  In our simple 
test cases this 
minimum usually occurs at $v_{sys}$.  However, in the 
\textit{BT} profile, this minimum is different than 
$v_{sys}$.  As such, we consider two versions of 
the asymmetry statistic.  The first is calculated 
at $v_{sys}$ that can be easily compared to the 
lopsidedness.  The second is calculated at 
the \bsq{velocity of symmetry},\textcolor{black}{$v_{sym}$}, (like the center of 
symmetry in 2-D images).  At this velocity, the 
asymmetry is minimized removing some of the 
uncertainty in the calculation of $\mathcal{A}$ that 
would arise due to uncertainties in calculating
$v_{sys}$.  Moreover, since the 
system itself may not be in equilibrium, the 
relation between a calculation of $v_{sys}$ and the true velocity
of the entire galaxy is not quite always clear.
In practical terms we find the velocity of 
symmetry using the \textsc{lmfit} \citep{LMFIT}
\textsc{python} package.  For simplicity we 
will use $\mathcal{A}$ for the 1-D asymmetry at the 
\bsq{velocity of symmetry} and $\mathcal{A}(v_{sys})$ 
when calculated at the systemic velocity.

\subsection{Dependence on Profile Signal-to-Noise Ratio}
\begin{figure*}
\centering
    \includegraphics[width=\textwidth]{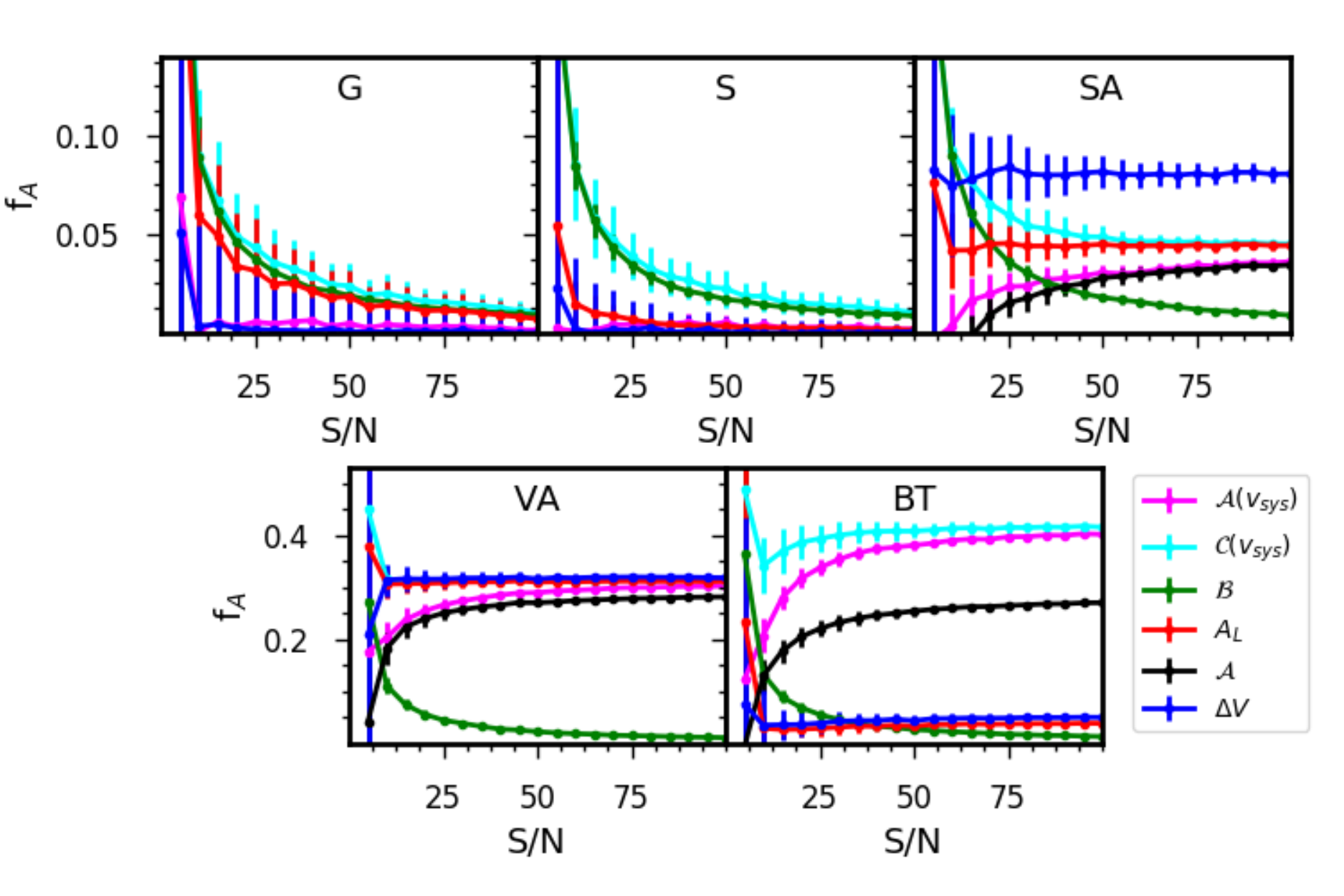} 
\caption{The asymmetry and lopsidedness statistics
for the five velocity profiles as a function
of $S/N$.  The magenta and red curves
are the asymmetry and lopsidedness measured at
$v_{sys}$.  The cyan and green curves are the 
average $\mathcal{C}$ and $\mathcal{B}$ terms
used to calculate $\mathcal{A}$.  The black
curve is the asymmetry statistic calculated at
the \bsq{velocity of symmetry}.  Finally, the blue 
curve is the $\Delta V$ statistic.  The 
points and error bars for each statistic 
were calculated from 100 realizations
of the velocity profiles at each $S/N$
ratio.}
  \label{Fig:NoiseResponse}
\end{figure*}

These profiles are also useful for exploring the 
effect of the signal-to-noise ratio, $S/N$, on the measured statistics.  We generated profiles with a specific
peak $S/N$ by first calculating the noise, $\sigma$.  The noise is found 
by dividing the peak flux by the target $S/N$.
Then a random value is drawn from a 
Gaussian distribution centered at zero with a dispersion
equal to $\sigma$ in all velocity channels and added to the profile.  In order to get robust 
results we generated 100 bootstrap samples
for each profile at each $S/N$ value.  Fig.
\ref{Fig:NoiseResponse} shows the average values
and 1-$\sigma$ error bars from those samples.

The effect of the noise on the lopsidedness is quite
interesting for two reasons.  Firstly, the value of $A_{L}$
decreases as a function of $S/N$ to some asymptotic value.
Secondly, and much more interestingly, the asymptotic value
is quite different for the \textit{VA} and \textit{BT} profiles.  While the 
two profiles appear to be nearly as asymmetric by eye,
the lopsidedness of the \textit{BT} profile is close
to zero.  The low lopsidedness
for the \textit{BT} profile is due to
the fact that the broad and narrow
peaks have nearly the same amount of
flux on either side of $v_{sys}$.  Therefore, while
the profile may appear to be asymmetric, the asymmetries
are local and there is no overall lopsidedness.

The $\Delta V$ statistic is more similar to the 
lopsidedness than the asymmetry. It is 
unsurprising that the two quantities are so similar, 
as $\Delta V$ depends on $A_{L}$.  However, in
the $SA$ case, $\Delta V$ is the largest statistic
for most $S/N$ values.  The key point here is that 
$\Delta V$ is not just a reformulation of $A_{L}$ and
gives different information about the profile than
the lopsidedness.

The channel-by-channel asymmetry exhibits a 
different behaviour.  Both $\mathcal{C}$ and
$\mathcal{B}$ decrease as a function of $S/N$, but the 
rate of decrease between these values differs.  
Because the decrease in $\mathcal{B}$ 
is smaller than $\mathcal{C}$, 
$\mathcal{A}$ increases to 
some asymptote.  Ultimately, the background
subtraction means that $\mathcal{A}$ is a measurement
of how much asymmetry may be robustly attributed
to the signal.  $\mathcal{C}$ includes
both contributions from the underlying signal and 
the noise.  In principle, $\mathcal{B}$ calculates
the amount of the asymmetry that could arise due
to random noise.  When the noise is large it can 
dominate over the asymmetry due to the signal.   
While there may be more \bsq{asymmetry} in the noise-free
profile, it is impossible to definitively attribute it
to the signal if the profile is noisy.

It is worth noting that in the \textit{BT} case 
$\mathcal{A}(v_{sys})$ is significantly higher than 
$\mathcal{A}$.  This makes sense as Fig. 
\ref{Fig:VSysResponse} shows that only in the 
\textit{BT} case is $v_{sym}$ significantly different than
$v_{sys}$.  This result highlights the importance of 
calculating the minimal asymmetry.

\textcolor{black}{
In summary, Fig. \ref{Fig:NoiseResponse} shows that
the lopsidedness and $\Delta V$ statistics reach an 
asymptotic value at lower S/N values than the
channel-by-channel asymmetry measurement.  However, the 
channel-by-channel asymmetry is still large for the visually
asymmetric \textit{BT} profile where the lopsidedness 
and $\Delta V$ are roughly zero.  This result emphasizes
that these different asymmetry statistics 
quantify different effects.  As such, when analyzing a
particular profile, all the statistics should be measured.
However, it is also important to account for biases that 
may arise due to the S/N when comparing different objects.
}

\subsection{Dependence on Profile Velocity Resolution}
\begin{figure*}
\centering
    \includegraphics[width=\textwidth]{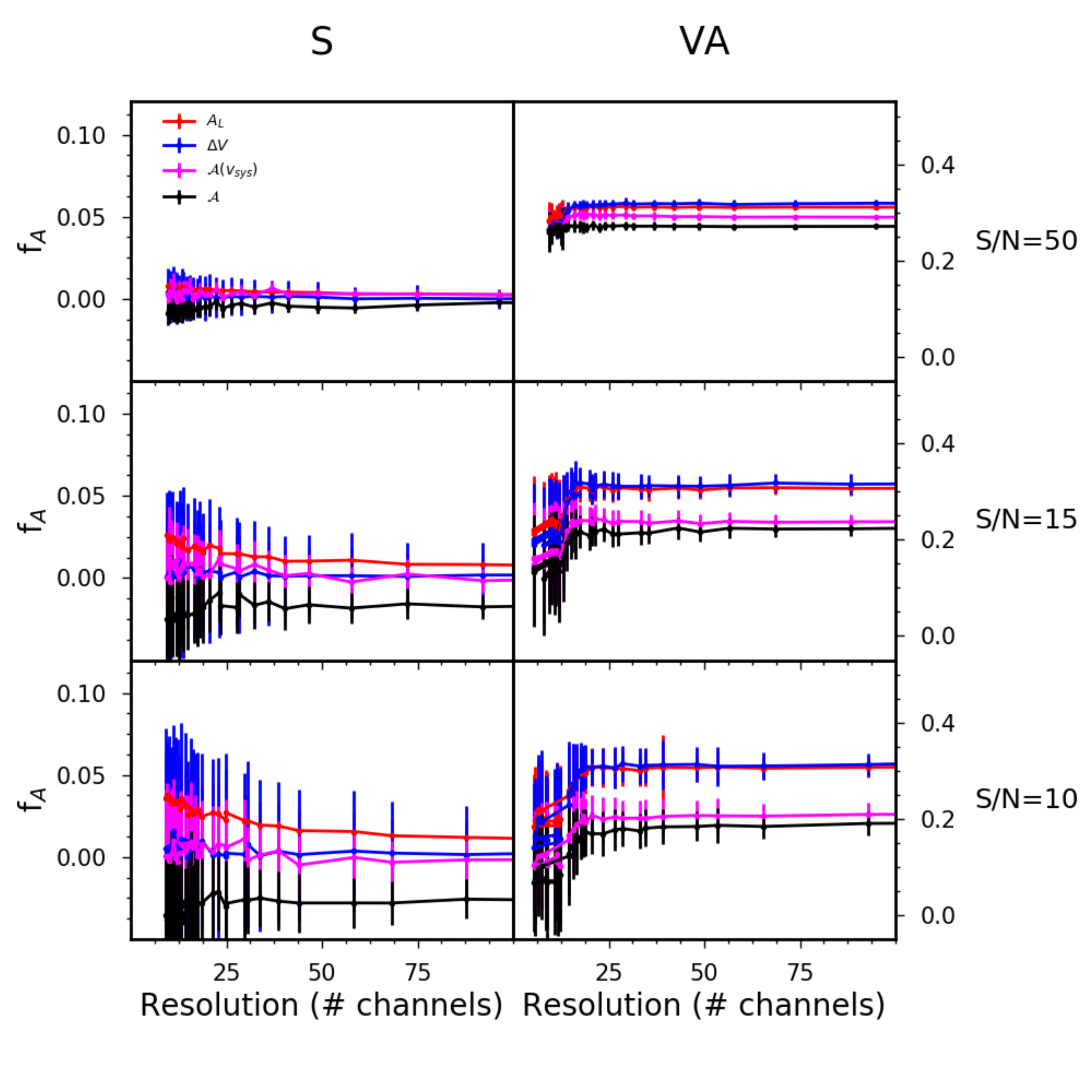} 
\caption{The resolution and S/N dependence for the
different asymmetry statistics.  The 
black, magenta, red, and blue lines are
the average $\mathcal{A}$, $\mathcal{A}(v_{sys})$,
$A_{L}$, and $\Delta V$ statistics for 100 bootstrap
samples respectively.
The resolution is the number of channels contained within
the profile.}
  \label{Fig:ResolutionDependence}
\end{figure*}

One potential issue with the channel-by-channel measures is
the dependence on the resolution of the \hi\ profile.  To that
end, Fig. \ref{Fig:ResolutionDependence} shows the 
average asymmetry statistics for a variety of 
resolutions and $S/N$ ratios.  For simplicity, 
this plot only shows the calculations for
the symmetric, $S$,  and very asymmetric,  $VA$, profiles.  As with the $S/N$ tests, we generated 100 realizations with its own random noise for 
each profile and plot the average values with dispersions indicated by the error bars on the data points.

Before discusssing the resolution effects, it's worth noting
that, for all $S/N$ values, $\mathcal{A}<0$ for
the $S$ profile.  This would also have been apparent in Fig.
\ref{Fig:NoiseResponse}, except the limits were set to 
zero in that plot.  We highlight this here to demonstrate
how symmetric or nearly symmetric profiles can
have $\mathcal{A}$ below zero.  As noted in 
Sec. \ref{SubSec:AsymmetryDef}, when the profile
is symmetric the background subtraction can reduce
a positive $\mathcal{C}$ below zero.

For both the $S$ and \textit{VA} profiles, a steady value for all the asymmetry parameters is reached for all $S/N$, for all resolutions above $\sim20$ channels. For profiles with fewer than 20 channels, the dispersions for all the asymmetry parameter measurements increase substantially as seen by the increase in the size of the error bars. Most notable is the effect on $\Delta V$ where the scatter is largest. 
For most of the asymmetry statistics, once the number of resolution elements drops below $\sim20$, the mean values of the measured statistics also drop, resulting in an under-estimate of the true asymmetry. An exception is the case for the lopsidedness, $A_{L}$, which actually increases in the case of a symmetric profile, therefore over-estimating the true asymmetry. In all cases, with increasing $S/N$, the effect of reduced resolution is minimized. In conclusion, the number of channels which an \hi\ profile spans is important to take into account when measuring the different asymmetry parameters, particularly when the $S/N$ of the profiles is low. For higher $S/N$ profiles, one can measure the asymmetry reliably at lower resolutions down to $\sim$15 channels. The parameter most susceptible to resolution effects is the $\Delta V$ statistic.

\section{Simulations}\label{Sec:Simulations}

In order to understand how the four asymmetry 
statistics depend on viewing angle and 
inclination, we utilize two simulations of 
merging galaxies.  The galaxies are 
generated using the \textsc{GalactICS} code
\citep{Deg2019} and the combined
systems are evolved using the Gadget-2 N-body code \citep{Springel2005}.  We then analyze the
profile generated from a single galaxy at a
variety of viewing angles and inclinations.

\subsection{Simulation Details}

We utilize the \textsc{GalactICS} code to generate
the initial conditions for the simulations.  The
full details of the code are described in 
\citet{Deg2019}.  In brief, \textsc{GalactICS}
generates equilibrium galaxy models
that may consist of a 
S\'{e}rsic bulge, a double-power law dark matter halo, 
up to two exponential-sech$^{2}$ stellar disks, and 
an exponential surface density gas disk.

We have run two simulations of merging galaxies in
order to explore the effects of viewing angle and
inclination on the various asymmetry statistics.
The first simulation, called the \bsq{Shock Sim} has 
the two galaxies crash into each other directly.
The second simulation, called the \bsq{Fly-by Sim} 
has the two merge over the course of a more 
gentle interaction.

\begin{figure}
    \centering
    \includegraphics[width=80mm]{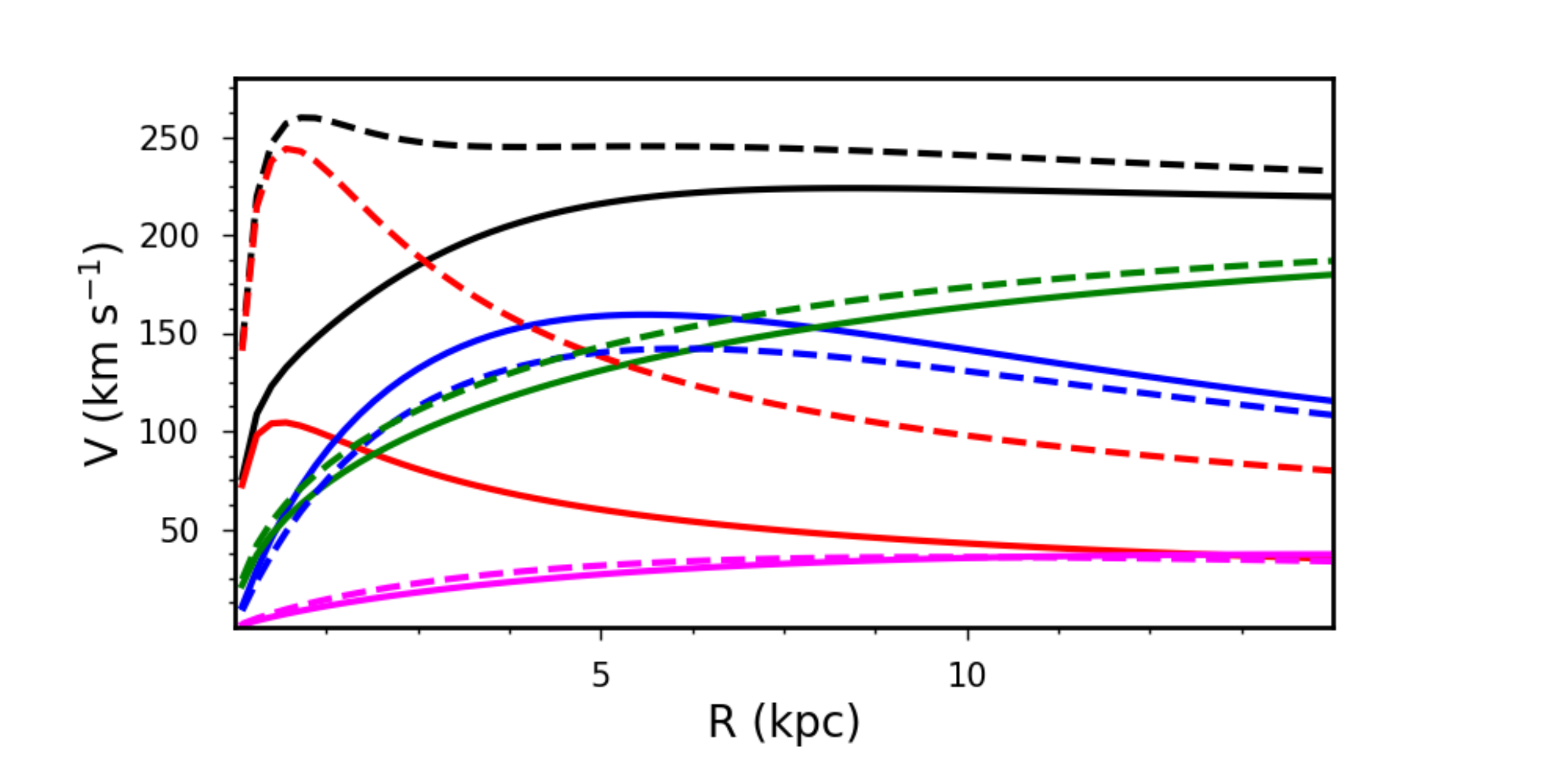}
    \caption{The rotation curves for the Shock Sim (dashed
    lines) and Fly-by Sim (solid lines).  The black lines
    are the total rotation curves, while the
    red, blue, magenta, and green lines are
    the bulge, stellar disk, gas disk, and halo
    contributions respectively.}
    \label{Fig:ModelRCS}
\end{figure}

Fig. \ref{Fig:ModelRCS} shows the initial rotation 
curves for each 
model.  Each galaxy in the Shock Sim has $10^5$ gas particles,
$10^{5}$ bulge particles, $2\times10^{5}$ disk particles,
and $10^{6}$ halo particles.  Each 
galaxy in the Fly-by Sim has $2\times 10^5$ gas particles,
$10^{5}$ bulge particles, $5\times 10^{5}$ disk particles,
and $10^{6}$ halo particles .

In both simulations, one galaxy
is placed at $(\mathbf{x},\mathbf{v})=(0,0)$.  For the 
Shock Sim, the second galaxy is initialized
at $\mathbf{x}=(0,300,0)$ kpc, 
$\mathbf{v}=(0,-25,0)$ km s$^{-1}$
There is no global rotation of the second
galaxy, making the interaction
edge-on and producing a strong shock
during the initial encounter. 
In the Fly-by Sim, the second 
galaxy begins at 
$\mathbf{x}=(85,-67,27)$ kpc, 
$\mathbf{v}=(64,320,-14)$ km s$^{-1}$
and is rotated so that the initial inclination
45$^{\circ}$ relative to the $X-Y$ plane.  This produces
a slower, fly-by first passage that then merges 
together over later passages.

Both systems are evolved using \textsc{Gadget-2}
for 5 Gyr.  The simulations use an adaptive
softening length of with 
a maximum length of 0.5 kpc.  The Courant factor used in
this simulation is 0.25.  In order to 
have significant time resolution, 
snapshots are produced every 0.005 Gyr.  

\begin{figure}
\centering
    \includegraphics[width=80mm]{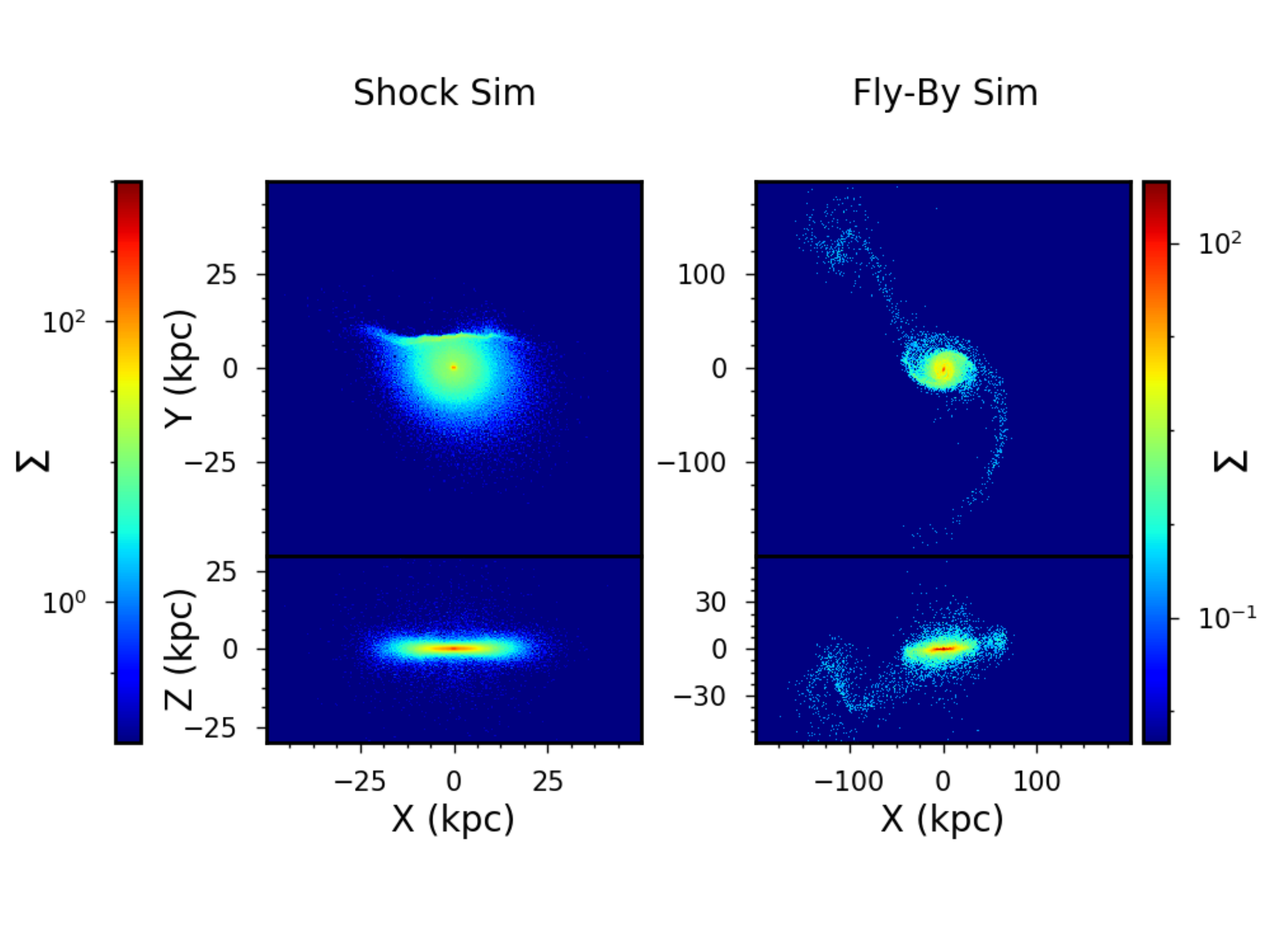} 
\caption{The gas disk surface density of the analyzed galaxy
for the Shock Sim (left) and Fly-by Sim (right).  The 
Shock Sim surface density is calculated at $T=1.93$ Gyr
and the Fly-by Sim is calculated at $T=2.5$ Gyr.
The surface
density units are $M_{\odot}$ pc$^{-2}$.}
  \label{Fig:SDComp}
\end{figure}

We have chosen a single snapshot from each simulation
for analysis.  For the Shock Sim, the snapshot is at 
$T=1.93$ Gyr, while for the Fly-by Sim we use the 
$T=2.5$ Gyr snapshot.  For simplicity we only 
analyze the gas particles that initially 
belong to a single galaxy in that snapshot.
Figure \ref{Fig:SDComp} shows the 
surface density maps selected for analysis. 
The Shock Sim shows a clear shock wave at the interaction
interface and the Fly-by Sim shows large tidal tails.  A consequence of this
is that some of the particles selected for analysis have
been captured by the second galaxy.  This selection is 
not realistic, but it is sufficient accurate for 
our focus on asymmetry statistics.  Future work will 
examine the effect of properly separating the different
galaxies from an observational point of view.

\subsection{Mock Observations}

We have generated mock observations of the 
Shock Sim and Fly-by Sim at a variety of viewing angles
and inclinations.  The selected particles from a 
particular snapshot are rotated using Euler angles, and 
then shifted to a new center.  Changing the Euler
angles changes the viewing angle and 
inclination of the galaxy.  A systemic 
velocity of $1000$ km s$^{-1}$ is then added to each galaxy.  
From this new position and velocity, $v_{r}$ is calculated
for an observer.  

To construct velocity profiles, we use 800 channels
with a velocity resolution of $5$ km s$^{-1}$.  The 
luminosity of each particle is convolved with 
a Gaussian profile with $\sigma=15$ km s$^{-1}$ and 
added to each channel.

\subsection{Viewing Angle}

Nature only gives us a single viewing angle and inclination 
for any particular interaction.  However, with a simulation, 
it is possible to vary these parameters and explore how they affect the 
measurement of some quantity of interest, in this case the 
asymmetry.

\begin{figure*}
\centering
    \includegraphics[width=150mm]{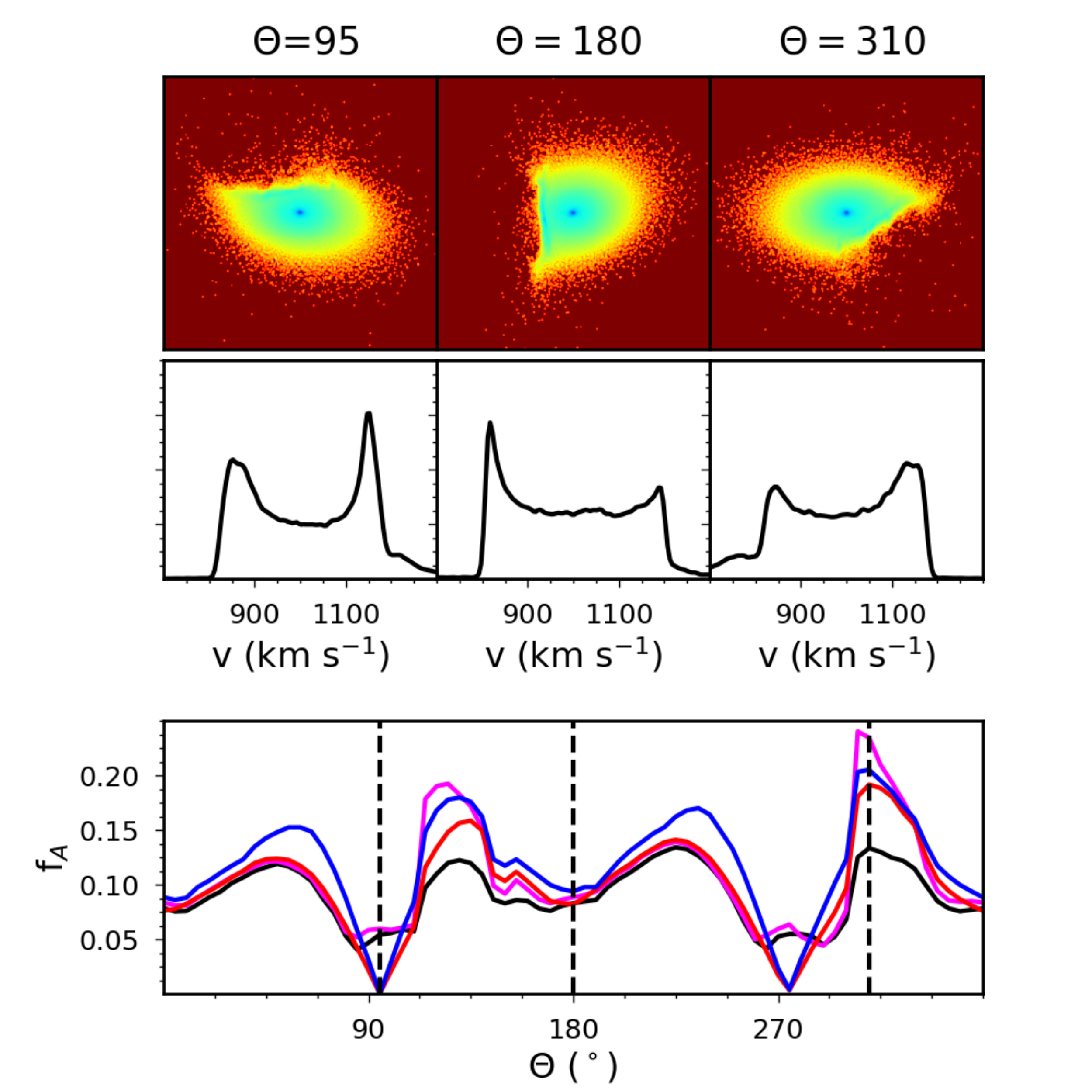}
\caption{The asymmetry statistics for the Shock Sim
as a function of viewing angle (bottom), along with a 
set of mock images (upper panels) and velocity profiles
(middle panels) at specific viewing angles.  The solid
black, red, blue and cyan lines are 
$\mathcal{A}$, $A_{L}$,
$\Delta V$, and $\mathcal{A}(v_{sys})$ respectively, while the dashed vertical lines
indicate the viewing angles of the images and velocity profiles shown in the panels above.}
  \label{Fig:ShockSimViewingAngle}
\end{figure*}

\begin{figure*}
\centering
    \includegraphics[width=150mm]{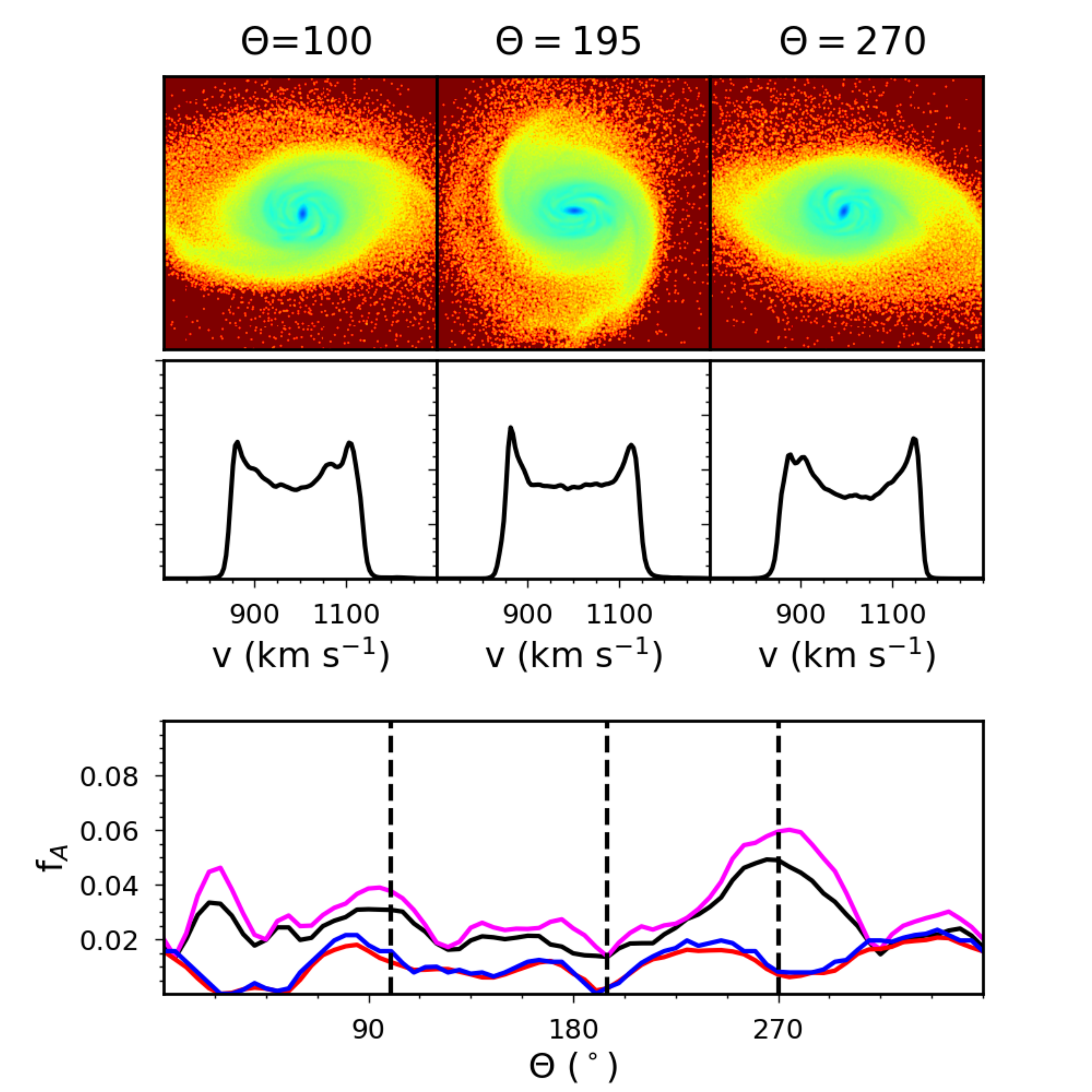}
\caption{The asymmetry statistics for the Fly-by Sim
as a function of viewing angle (bottom), along with a 
set of mock images (upper panels) and velocity profiles
(middle panels) at specific viewing angles.  The lines are
the same as in Fig. \ref{Fig:ShockSimViewingAngle}. }
  \label{Fig:FlyBySimViewingAngle}
\end{figure*}

Figure \ref{Fig:ShockSimViewingAngle} shows the dependence of 
the four asymmetry statistics $\mathcal{A}$, $A_{L}$,
$\Delta V$, and
$\mathcal{A}(v_{sys})$ as a function of viewing
angle for the Shock Sim.
Figure \ref{Fig:FlyBySimViewingAngle} shows the 
same quantities for the Fly-by Sim.  For 
each of these plots, the observed galaxy has
an inclination of $45^{\circ}$.  It is worth 
noting that the sample images
in Fig. \ref{Fig:FlyBySimViewingAngle} do not show the 
full extent of the tidal tails.  However, the velocity 
profile includes the contribution of all the 
gas particles, whether they are in the main body, the 
tidal tail, or trapped in the second galaxy.

Firstly, the 
viewing angle affects all four asymmetry 
statistics.  This is 
due to the fact that velocity profiles combine both 
kinematic and morphological information.  The 
images and velocity profiles shown in Fig. 
\ref{Fig:ShockSimViewingAngle} demonstrate this 
quite clearly.  The image is approximately the same for
each of the selected viewing angles, but the 
projection of the shock towards the observer changes.  This
projection causes the large variations in the observed
velocity profiles.

The second thing to note in the two figures 
is that $\mathcal{A}$
seems to be least sensitive of the 
statistics to viewing 
angle variations.  This is especially clear in 
Fig. \ref{Fig:ShockSimViewingAngle}, where, for certain viewing
angles like $\Theta=95^{\circ}$, $A=\Delta V=0$.  
That specific profile is similar to our 
test \bsq{broad-thin} profile, and the
flux on either side of the systemic velocity is the same.
At other angles, such as $\Theta=290^{\circ}$, the 
lopsidedness and $\Delta V$ statistics are larger
than $\mathcal{A}$.  The first takeaway from this is that
the viewing angle has a strong effect on the calculation
of any asymmetry statistic.  In some cases it is 
possible for a very asymmetric galaxy image to have 
profiles with no global lopsidedness.  
\textcolor{black}{Thus, while \textcolor{black}{a} lopsided profile
almost certainly indicates some disturbance to the galaxy, 
low lopsidedness does not indicate an undisturbed galaxy.}
The 
second point is that $\mathcal{A}$ is the most stable
of the asymmetry statistics and it never indicates that
the profile is symmetric due to its sensitivity to 
local perturbations in the profile.

\textcolor{black}{A third interesting, but somewhat more 
minor, result is the sensitivity of the three statistics
to the low flux \bsq{wings} of the Shock-Sim profile.
As the viewing angle changes a low flux 
feature moves through the channels.  This feature is 
seen in the high velocity wing of the  $\Theta=95^{\circ}$
profile
and at the low velocity wing of the $\Theta=310^{\circ}$
profile.  This feature is actually due to the projection
of the shock velocity along the line-of-sight.  The effect
of this feature, and in particular, the wings, is somewhat
subtle and may suggest another diagnostic for 
future work.  When the shock wave creates a profile wing
it changes the estimate of $v_{sys}$ due to 
our definition of the profile
edges being the velocities that have
20\% of the peak fluxes.  The wings have enough flux at 
certain orientations to move the edge values and thereby 
change $v_{sys}$.  This in turn strongly affects $A_{L}$,
$\Delta V$, and $\mathcal{A}(v_{sys})$.  It has a much
smaller effect on $\mathcal{A}$ as that is calculated 
at $v_{sym}$.  In Fig. \ref{Fig:ShockSimViewingAngle}
the orientations where the wings are included or 
skipped can be clearly seen by the jumps in 
$A_{L}$, $\Delta V$, and $\mathcal{A}(v_{sys})$
at $\Theta= 115^{\circ}$ and $\Theta=300^{\circ}$.
In this work we have restricted ourselves to the 
20\% edge definition.  At
the 50\% limits, the low flux wings would not affect the 
edge locations.  As such, $v_{sys}$ would be more stable
with viewing angle and the shape jumps seen in the asymmetry 
statistics would not be seen.  This suggests that a 
comparison of $v_{sys}$, $A_{L}$, $\Delta V$, or 
$\mathcal{A}(v_{sys})$ calculated at both the 20\% and 50\% 
limits might be used to detect global profile wings.  
However, such a diagnostic would have to be used carefully
due to the effects of noise in more realistic measurements.
This possibility is something that we will explore
in future work.
}

\subsection{Inclination}

\begin{figure*}
\centering
    \includegraphics[width=150mm]{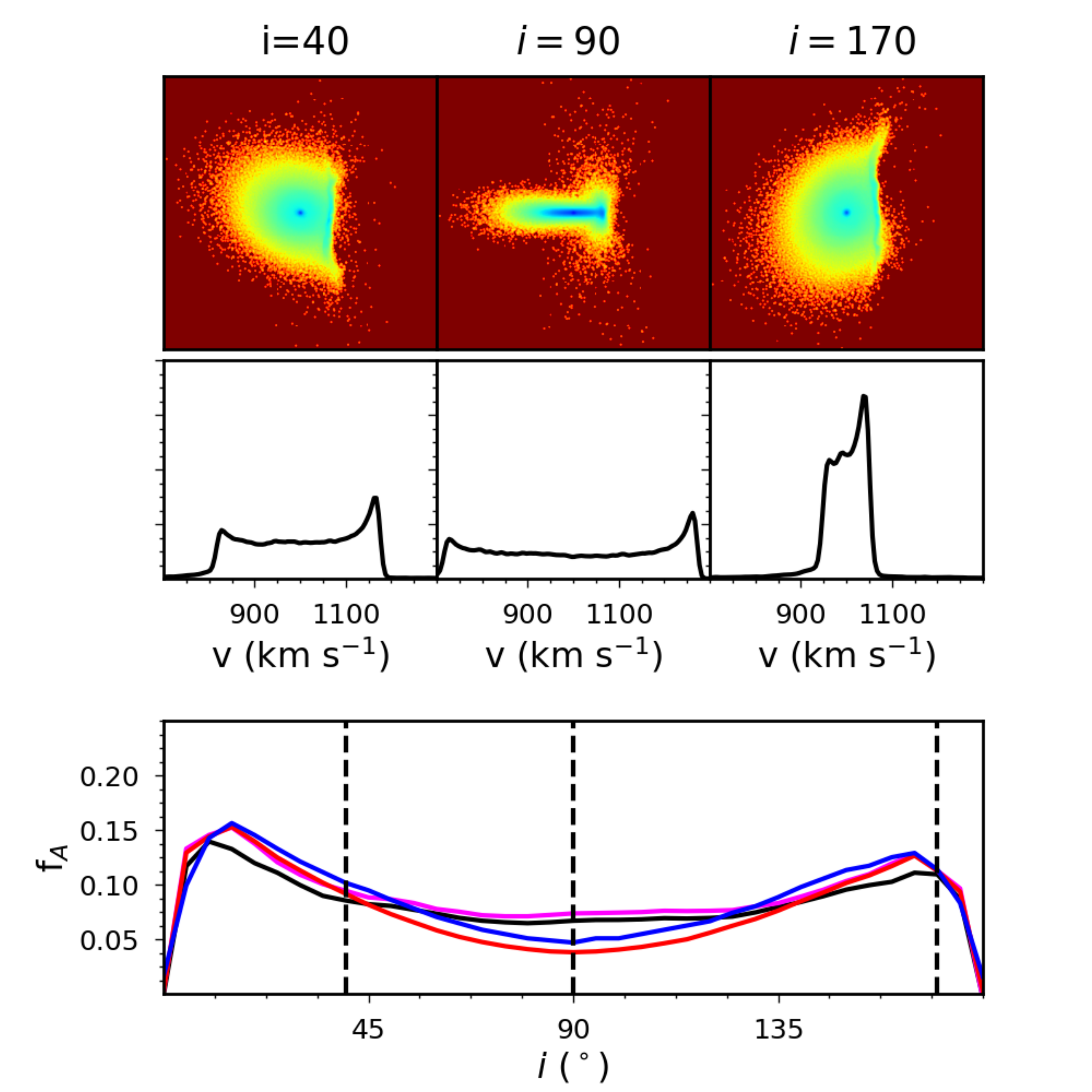}
\caption{The asymmetry statistics for the Shock Sim
as a function of inclination (bottom), along with a 
set of mock images (upper panels) and velocity profiles
(middle panels) at specific inclinations.  The lines are
the same as in Fig. \ref{Fig:ShockSimViewingAngle}.}
  \label{Fig:ShockSimInclination}
\end{figure*}

\begin{figure*}
\centering
    \includegraphics[width=150mm]{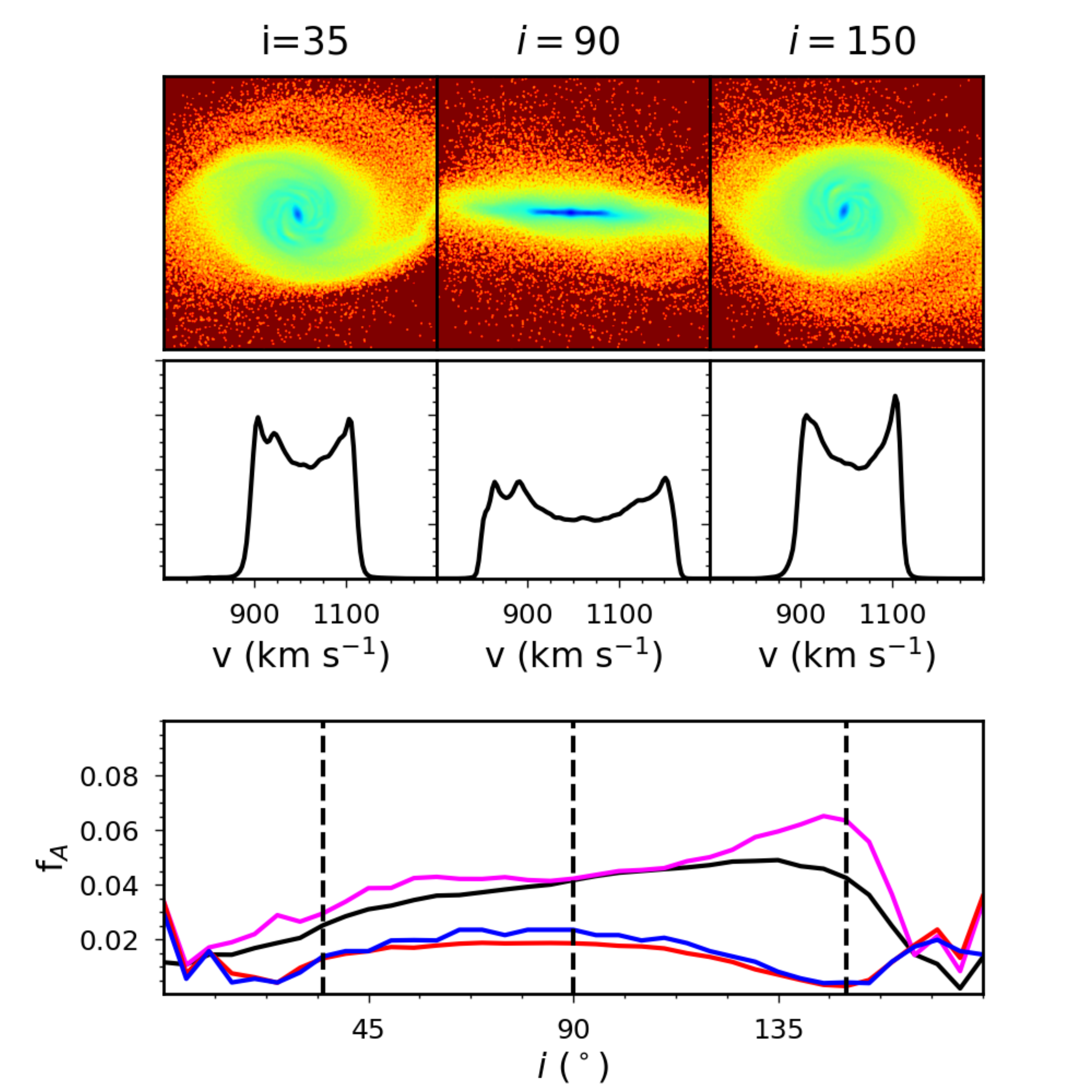}
\caption{The asymmetry statistics for the Fly-by Sim
as a function of inclination (bottom), along with a 
set of mock images (upper panels) and velocity profiles
(middle panels) at specific inclinations.  The lines are
the same as in Fig. \ref{Fig:ShockSimViewingAngle}. }
  \label{Fig:FlyBySimInclination}
\end{figure*}

The inclination of the observed galaxy must also 
have an effect on the asymmetry statistics.  To 
explore this, Figs. \ref{Fig:ShockSimInclination}
-\ref{Fig:FlyBySimInclination} show
the inclination dependence for the Shock Sim and
Fly-by Sim respectively.  In 
Fig. \ref{Fig:ShockSimInclination} the Shock-Sim is 
viewed at $\Theta=0^{\circ}$, while in 
Fig. \ref{Fig:FlyBySimInclination} the Fly-by Sim is 
viewed at $\Theta=290^{\circ}$.  The viewing angles
are selected to have enough asymmetries in the 
profile at moderate inclinations that they can 
show the effect of inclination variation.
Additionally, it is worth noting that the profiles have
very high $S/N$ ratios and the bin widths are 
$5~\kms$.  As such, the decrease in number of channels
in near face-on galaxies will still be in the reliable
regime according to Fig. \ref{Fig:ResolutionDependence}.

The inclination dependence of the Shock Sim is quite interesting.
Both $\mathcal{A}$ and 
$\mathcal{A}(v_{sys})$ are mostly constant along the 
range of inclinations, while the
lopsidedness and $\Delta V$ generally increase as the system
becomes more face-on.  
Once the galaxies become nearly face-on, all the 
asymmetry statistics go to zero.  The collapse at this orientation is due
to the profile reducing to a single-peaked Gaussian.  Since
the impact is edge-on, there are little, if any, systemic
$z$ velocities that project towards the observer.  

In the Fly-By Sim, the vertical motions do 
cause changes in channel-by-channel statistics.  Most
interestingly, there is a peak in $\mathcal{A}$ near
$\sim 135^\circ$ and in $\mathcal{A}(v_{sys})$ 
at $\sim 150^\circ$.  This is due to how the vertical 
perturbations of the galaxy project into the profile.  These
are also very local perturbations and give little to 
no lopsidedness.

\textcolor{black}{The combination of 
Figs. \ref{Fig:ShockSimInclination}-
\ref{Fig:FlyBySimInclination} suggest that asymmetry 
measurements are relatively robust for inclinations 
greater than 15-20$^\circ$.  When galaxies are more face-on
than this, the asymmetries decrease rapidly as the profiles
become very narrow.  Related to this 
is the requirement of at least 10 resolution elements for 
even high S/N profiles (based on Fig.
\ref{Fig:ResolutionDependence}.)}

\section{The WHISP Sample}\label{Sec:WHISP}

Moving from simulations, we have applied our statistics
to a subsample of WHISP galaxies.  The goal of this
work is to explore how the various asymmetry statistics
correlate with features of velocity profiles.  To that 
end we also visually classified each profile
using a variety of features (number of peaks, central
dip depth, apparent asymmetry, etc.).  

\subsection{Data}
We used observational data from
WHISP \citep{vanderHulst2001} to generate a subsample
of velocity profiles for analysis.
WHISP is an interferometric survey of 375 mostly late-type galaxies using the Westerbork Synthesis Radio Telescope and the data cubes, \hi\ images, and velocity fields are publicly available\footnote{http://wow.astron.nl/}. Three versions of the data cubes are available processed with different spatial resolutions.  We chose to use the intermediate cubes with 30 arcsec $\times$ 30 arcsec resolution so as to optimise the the signal-to-noise ratio per pixel while retaining good spatial resolution needed for two dimensional asymmetry analysis to be published in an upcoming paper, \citet{Hank2020}. 
For most of the galaxies in the sample, the velocity resolution is 5 km~s$^-1$.

Rather than using the full sample of WHISP galaxies,
we have restricted ourselves to a subsample consisting
of galaxies defined in \citet{Swaters2002} and
\citet{Noordermeer2005}.  These galaxies have 
good $S/N$ ratios, spatial and spectral resolution, and 
are sufficient for both the 1D analysis performed
here and the 2D work of \citet{Hank2020}.  They have
also been well-studied in a variety of 2D morphological
studies, including \citet{Holwerda2011} and \citet{Giese2016}.
Since we did not have the original mask cubes, we extracted the velocity profiles by using the \hi\ image of each galaxy as a mask. \textcolor{black}{We removed negative pixels, assuming they were due to noise, by calculating the RMS of only negative pixels in the image and then performing a 1$\sigma$ cut on the full image before applying it as a mask.}
We applied the same mask region to each channel of the cube for each galaxy respectively and summed the total flux density in each channel to create a spectrum. As a result of not optimising the mask for each channel individually, the extracted profiles typically consisted of \hi\ signal on a background pedestal. We estimated the pedestal by fitting a straight line between the average flux density values to the left and right away from the \hi\ line and subtracted it from the spectrum to extract the final \hi\ line profile.  Some of the profiles
did not include enough channels to properly do a pedestal
subtraction, leaving us with a final sample of 116 galaxies. 
The average peak $S/N$ of these profiles in $\sim30$ and 
all galaxies have sufficient channels for reliable 
asymmetry calculations.

\subsection{By Eye Classification}

Each velocity profile is both analyzed using the 
asymmetry statistics and classified by eye.  The purpose of the 
by-eye classification is to see how well $\mathcal{A}$, $A_{L}$, or 
$\Delta V$ correlate with our judgement of the level of 
asymmetry in \textcolor{black}{a global HI profile}.  Moreover, by classifying the
galaxies, we can look for any clustering in parameter space and
determine if there are any correlations.

Our visual classification scheme is quite basic.  Firstly, we decide if 
a profile has a single or double peak.  We then determine
whether the profile is symmetric, slightly asymmetric, 
moderately asymmetric, or very asymmetric.  We examine whether 
the central region of the profile dips below 50\% of the peak
flux.  This can only be true for double-peaked profiles.  Finally,
we classify the relative peak widths.  This classification is the 
most complicated as there are three categories.  The profile may 
have two peaks of differing widths, a single peak that is 
\bsq{diagonal} (having strongly differing slopes leading to the peak),
or the two peaks/one peak are the same width/symmetric respectively.

Three of the authors classified the galaxies first individually.  Then, in 
cases where at least one person
determined different levels of asymmetry, the three classifiers re-examined
and re-classified those profiles together to obtain a consensus classification for all galaxies.

\subsection{WHISP Asymmetries}

\begin{figure*}
\centering
    \includegraphics[width=\textwidth]{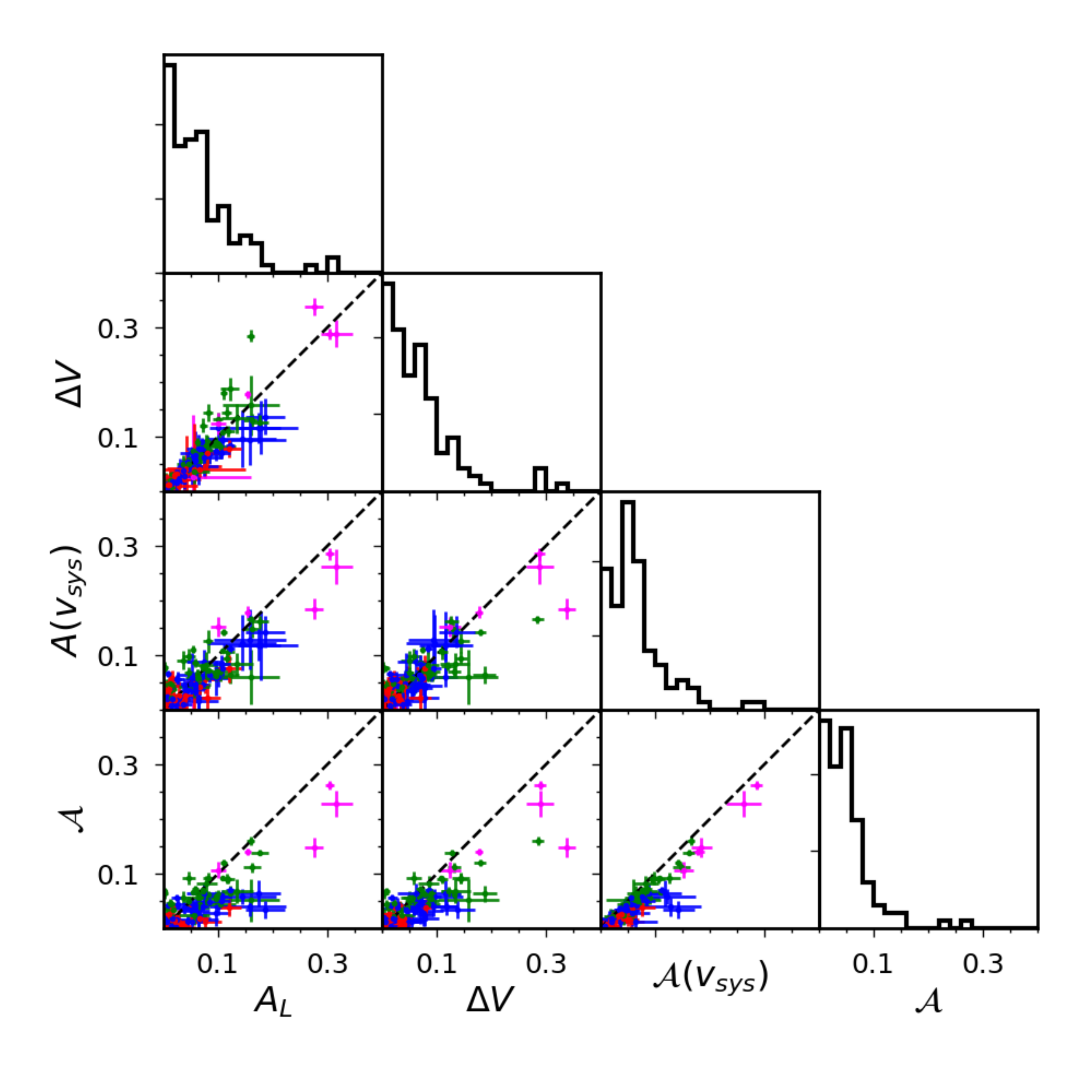}
\caption{The WHISP galaxy asymmetry statistics.  The red, blue, green, and \textcolor{black}{magenta} points are profiles classified as symmetric, 
slightly asymmetric, moderately asymmetric, and very asymmetric
by eye.  The dashed black lines in the correlation 
plots shows the 1-1 line.  The histograms show the distribution
of each statistic for the full sample.  The error bars were 
determined by creating 100 mock profiles with Gaussian noise 
added to them based on the measured $S/N$.  The one sigma
error bars calculated from that sample are shown in the 
correlation plots.}
  \label{Fig:AsymCorr}
\end{figure*}

Figure \ref{Fig:AsymCorr} shows the correlations between the 
three asymmetry statistics for our sample of
116 galaxies.  There are a number of things to note in 
this plot.  Firstly it is clear that $A_{L}$ and 
$\Delta V$ are strongly correlated.  This agrees with the 
results from the viewing angle and inclination studies.
Given that $\Delta V$ is strongly related to the 
lopsidedness, this result is perhaps not too surprising.  

There is a much weaker correlation of either
channel-by-channel asymmetry measurement with 
the lopsidedness or $\Delta V$.  This is not surprising as
$\mathcal{A}$ is more sensitive to local asymmetries than either
of the other statistics.  There is a correlation between
$\mathcal{A}$ and $\mathcal{A}(v_{sys})$, where
$\mathcal{A}$ is, by design, always smaller.
Nonetheless, this linear plot makes it somewhat difficult to 
investigate what is occurring at lower asymmetry levels.

It is also worth noting that the error bars on $\mathcal{A}$
are generally smaller than those on $A_{L}$, $\Delta V$, or
even $\mathcal{A}(v_{sys})$.  This is
due to the fact that the asymmetry statistic is measured at 
\textcolor{black}{$v_{sym}$} and does not depend on $v_{sys}$.  Therefore errors in 
the calculation of $v_{sys}$ due to noise will not propagate into
$\mathcal{A}$.  

\begin{figure*}
\centering
    \includegraphics[width=\textwidth]{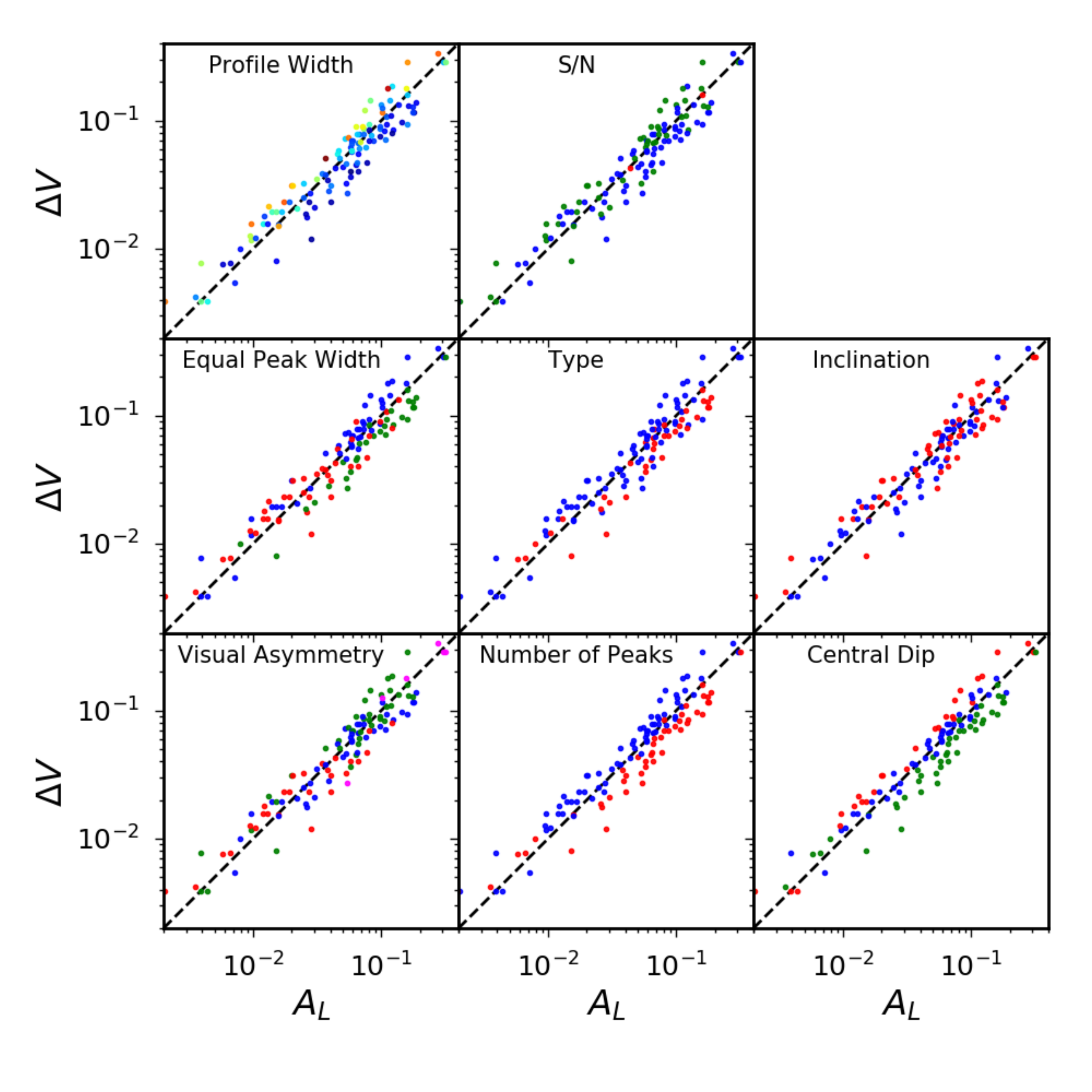}
\caption{Lopsidedness-$\Delta V$ correlations.  The bottom-left
panel uses the same colors as Fig. \ref{Fig:AsymCorr}. In the 
bottom- middle panel the red and blue points have one or 
two peaks.  In the bottom-right panel the green, blue, and 
red points have a single peak, a central dip that is 
above the 
50\% flux level, and a central dip below the 50\% level
respectively.  In the middle-left panel the green, blue, and 
red points have single peaks with differing slopes, two peaks
with differing widths, or have symmetrical peak widths/single
peaks that have similar slopes.  In the middle panel, the 
red and blue points are irregular or spiral galaxies. 
\textcolor{black}{The middle right panel has galaxies with
$30^\circ \le i \le 60^{\circ} $ in blue and the other 
inclinations in red}.  The 
upper-left panel has a continuum of colors where dark blue 
points have small profile widths and cyan-red points have
larger profile widths, with red as the largest. 
\textcolor{black}{Finally,
the upper-middle panel has objects with S/N$<10$ in red, 
those with $10 \le $S/N$ < 30$ in blue and those with S/N $\ge 30$ in green.}}
  \label{Fig:LogDVLopCorr}
\end{figure*}

In order to examine some of the correlations and possible
dependencies in greater detail, Fig. \ref{Fig:LogDVLopCorr}
shows the $A_{L}$-$\Delta V$ correlations on a log-log
scale.  In general, all the WHISP galaxies lie along the 
1-1 line.  This is unsurprising given that $\Delta V$ is
related to $A_{L}$ by $v_{equal}$.

There are a few relations that become clear in this figure.
Firstly, and perhaps most importantly, the visual asymmetry
classification spans the parameter space of both $A_{L}$ and 
$\Delta V$.  There are galaxies classified as asymmetric with fairly
low values of $A_{L}$ and $\Delta V$. This is due to the 
fact that
 the \bsq{by-eye} classification
accounts for both local and global features.

Secondly, there are a number of offsets 
that appear in different
panels in Fig. \ref{Fig:LogDVLopCorr}.  For instance, 
the irregulars are generally separated from the spirals in 
the middle panel.  A clearer separation appears between 
the single and double peaked profiles.  The cause of this
separation, as well as the \bsq{equal peak width} separation is 
ultimately due to the trends seen in the \bsq{central-dip} panel.

\begin{figure}
    \centering
    \includegraphics[width=80mm]{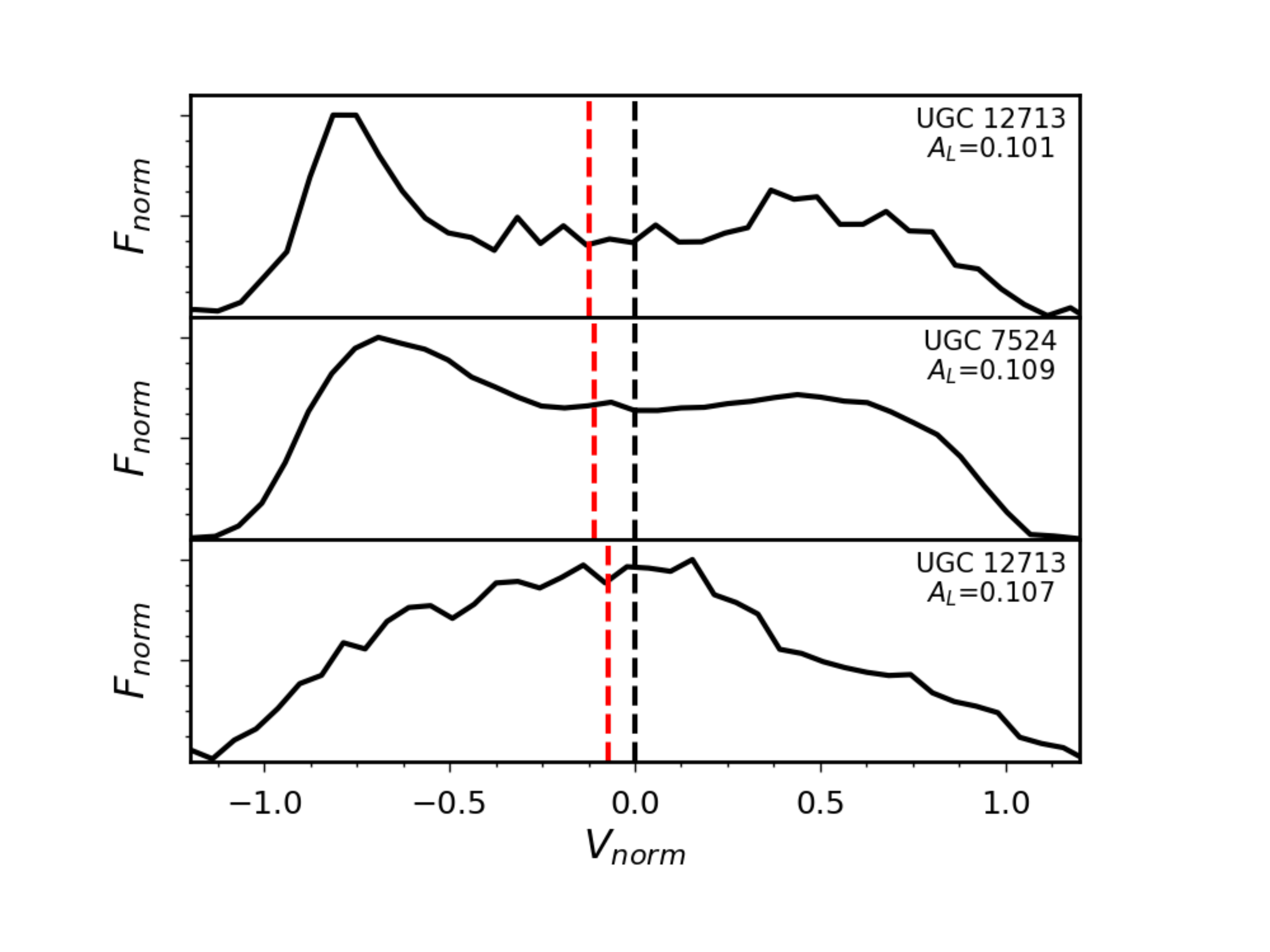}
    \caption{Three sample profiles with similar 
    values of $A_{L}$.  The y-axis is the flux normalized 
    by the peak flux of each profile, while the x-axis shows 
    the velocity shifted by $v_{sys}$ and normalized by the 
    width of each profile.  The solid black line shows
    the profile, while the vertical dashed black and red lines
    show $v_{sys}$ and the value of $v_{fold}$ where the integrated
    flux on either side of the red line is equal.}
    \label{Fig:DeltaVExamples}
\end{figure}

To understand the central dip relationships Fig. 
\ref{Fig:DeltaVExamples} shows three profiles with 
similar values of $A_{L}$.  This figure shows that, 
for given values of $A_{L}$, $\Delta V$ tends to be higher
 when the central dip is deeper (see the horizontal difference
 between the red and black vertical dashed lines in
 Fig. \ref{Fig:DeltaVExamples}).  This is due to the 
 nature of the $\Delta V$ statistic itself.  As
 $v_{fold}$ moves away from $v_{sys}$ it moves flux 
 from one of the integrals to the other.  
 This increases/decreases $L$ and inversely affects $H$.
 If $v_{sys}$ is located 
 near a peak, shifts in $v_{fold}$ will move proportionally
 more flux than cases where $v_{sys}$ is located near the 
 central dip.  As such, profiles with deeper dips will need
 to move $v_{fold}$ further away to get equivalent shifts of
 flux from $L$ to $H$.  Therefore, for a given 
 value of $A_{L}$, $\Delta V$ will be larger when the 
 central dip is deeper.

This result also explains why profiles with greater 
widths generally lie above the one-to-one line. Wider 
profiles are more often double-peaked with 
deeper dips.  Inversely, inclination can 
shrink profiles with deep dips, and, in the 
limit of face-on, transform them to single
peaked profiles.  Therefore the narrow profiles 
will have fewer deep central dips and lie
lower on the $A_{L}-\Delta V$ trend.

\textcolor{black}{In this figure there is no clear
dependence on inclination.  However there is a hint
of a trend in the S/N, with higher S/N objects having
larger $\Delta V$ values.  It is also worth noting
that the WHISP sample is selected to have both 
high peak S/N values, thus there are few objects below a
S/N cut of 10.}

\begin{figure*}
\centering
    \includegraphics[width=\textwidth]{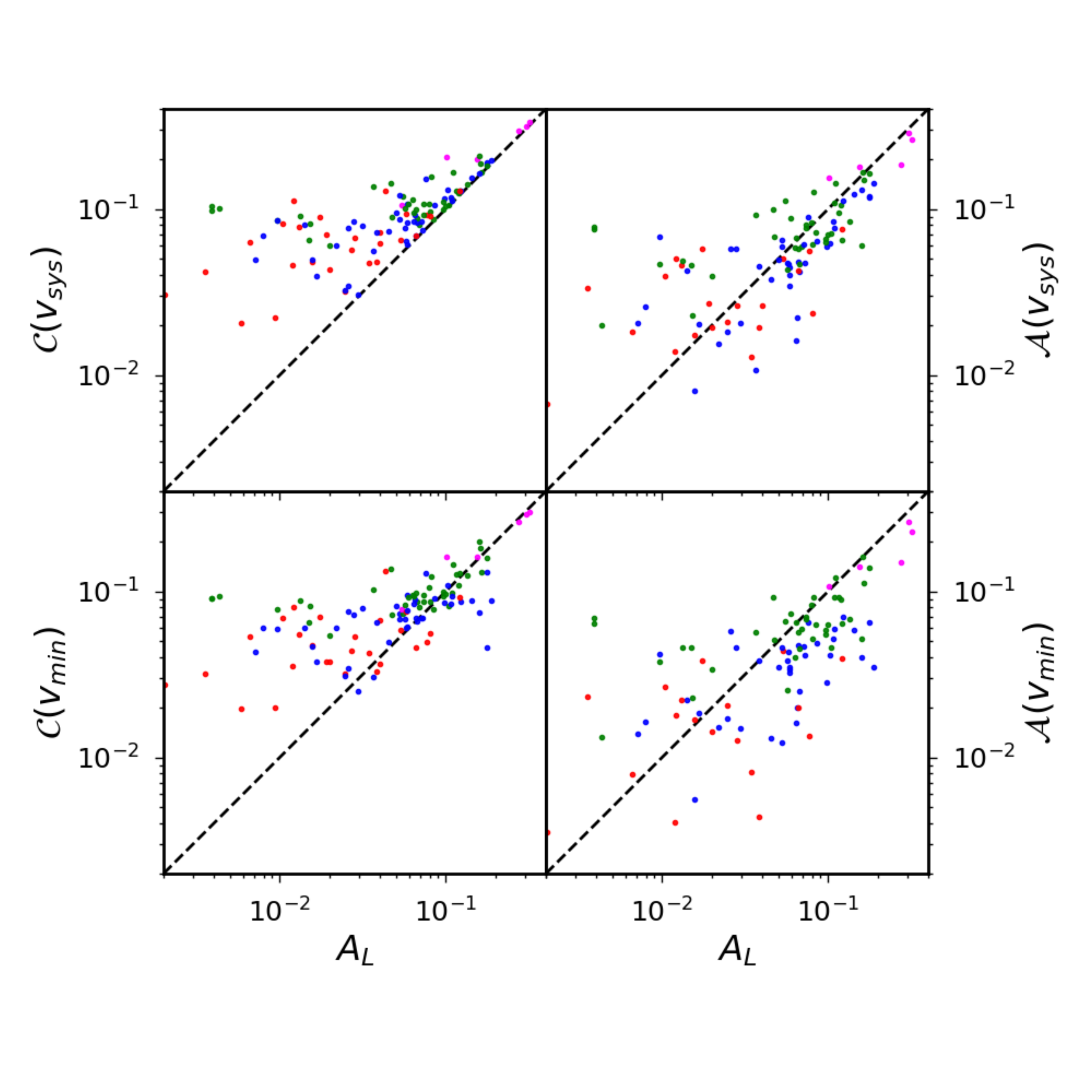}
\caption{Detailed Asymmetry-Lopsidedness correlations.  The
point colors represent our
visual classification for each galaxy and 
match the colors in \textcolor{black}{ Fig.
\ref{Fig:AsymCorr}}.  The upper-left panel shows
the correlation between $\mathcal{C}$ calculated
at $v_{sys}$ and $A_{L}$ while the lower-left panel
shows $\mathcal{C}$ calculated at \textcolor{black}{$v_{sym}$}.
The upper-right and lower right panels show
$\mathcal{A}$ calculated at $v_{sys}$ and \textcolor{black}{$v_{sym}$}
respectively.}
  \label{Fig:DetailedLogAsymLopCorr}
\end{figure*}

Moving from the $A_{L}-\Delta V$ correlations, 
Fig. \ref{Fig:DetailedLogAsymLopCorr} shows
the difference
between the signal asymmetry, $\mathcal{C}$ and the measured
asymmetry, $\mathcal{A}$, and how they 
correlate with $A_{L}$.  When calculated at $v_{sys}$, 
$\mathcal{C}$ is always larger than $A_{L}$.  This
naturally flows from the definitions of
Eq. \ref{Eq:NewLopsidedness} and 
Eq. \ref{Eq:1D_Asymmetry_NoBack}.  When $\mathcal{C}$ is calculated
at the \bsq{velocity of symmetry}, $\mathcal{C}$ decreases.
One of the effects of this decrease is to help to 
separate the profiles by their visual classification.
However, the greatest separation of the profiles
by visual classification is in $\mathcal{A}$ calculated
at \textcolor{black}{$v_{sym}$}.  In that panel, it is clear that 
the asymmetry statistic, calculated at the 
\bsq{velocity of symmetry}, with the background
subtraction, is the most correlated with our 
visual asymmetry classification. 

\begin{figure*}
\centering
    \includegraphics[width=\textwidth]{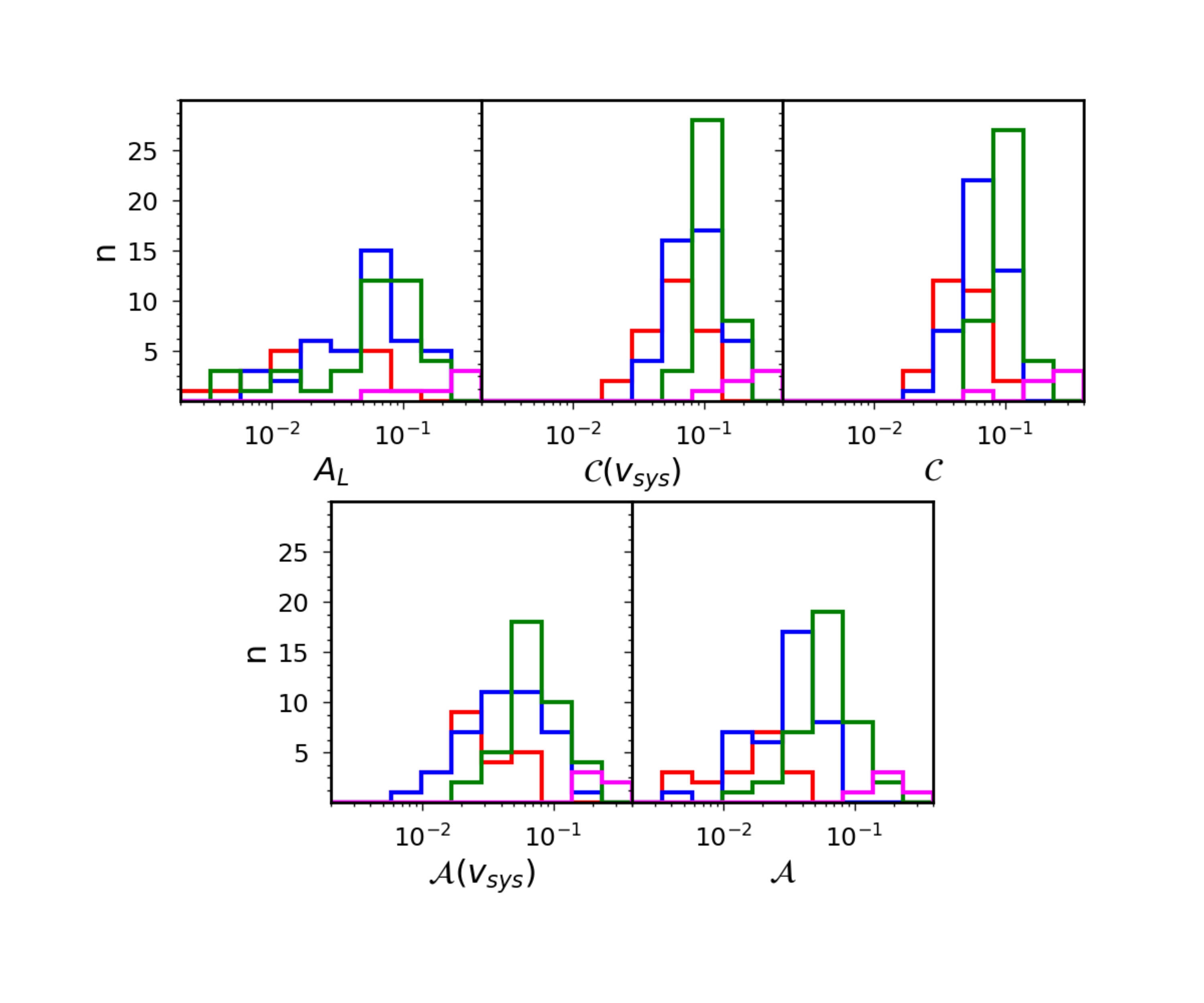}
\caption{\textcolor{black}{Histograms of the different
asymmetry statistics depending on the visual classification.
In each panel the \textcolor{black}{red, blue, green and magenta lines}
represent profiles that are visually classified as 
symmetric, slightly asymmetric, moderately asymmetric, 
and very asymmetric \textcolor{black}{(as in Fig. \ref{Fig:AsymCorr})}.  The upper-left panel shows the 
lopsidedness, the upper-middle panel shows $\mathcal{C}$
calculated at $v_{sys}$ and the upper-right 
panel shows $\mathcal{C}$ calculated at $v_{sym}$.
The lower-left and lower-right panels show 
$\mathcal{A}$ calculated at $v_{sys}$ and $v_{sym}$
respectively.}}
  \label{Fig:HistogramComparison}
\end{figure*}

\textcolor{black}{To help make
this point more clear, Fig. \ref{Fig:HistogramComparison}
shows histograms of these 5 statistics for each of the 
visual classifications.  The large overlap in histograms
for $A_{L}$ indicates that the visual classifications are 
not strongly correlated with lopsidedness.  
$\mathcal{C}$ is reasonably
separated, but, since there is no background subtraction,
the symmetric, slightly asymmetric, and moderately 
asymmetric still have large amounts of overlap.  The 
greatest separation of the different histograms
is for $\mathcal{A}$.  This indicates that the 
full channel-by-channel asymmetry, measured at the 
\bsq{velocity of symmetry}, is the asymmetry statistic that 
correlates best with visual asymmetry classifications.}  

\begin{figure*}
\centering
    \includegraphics[width=\textwidth]{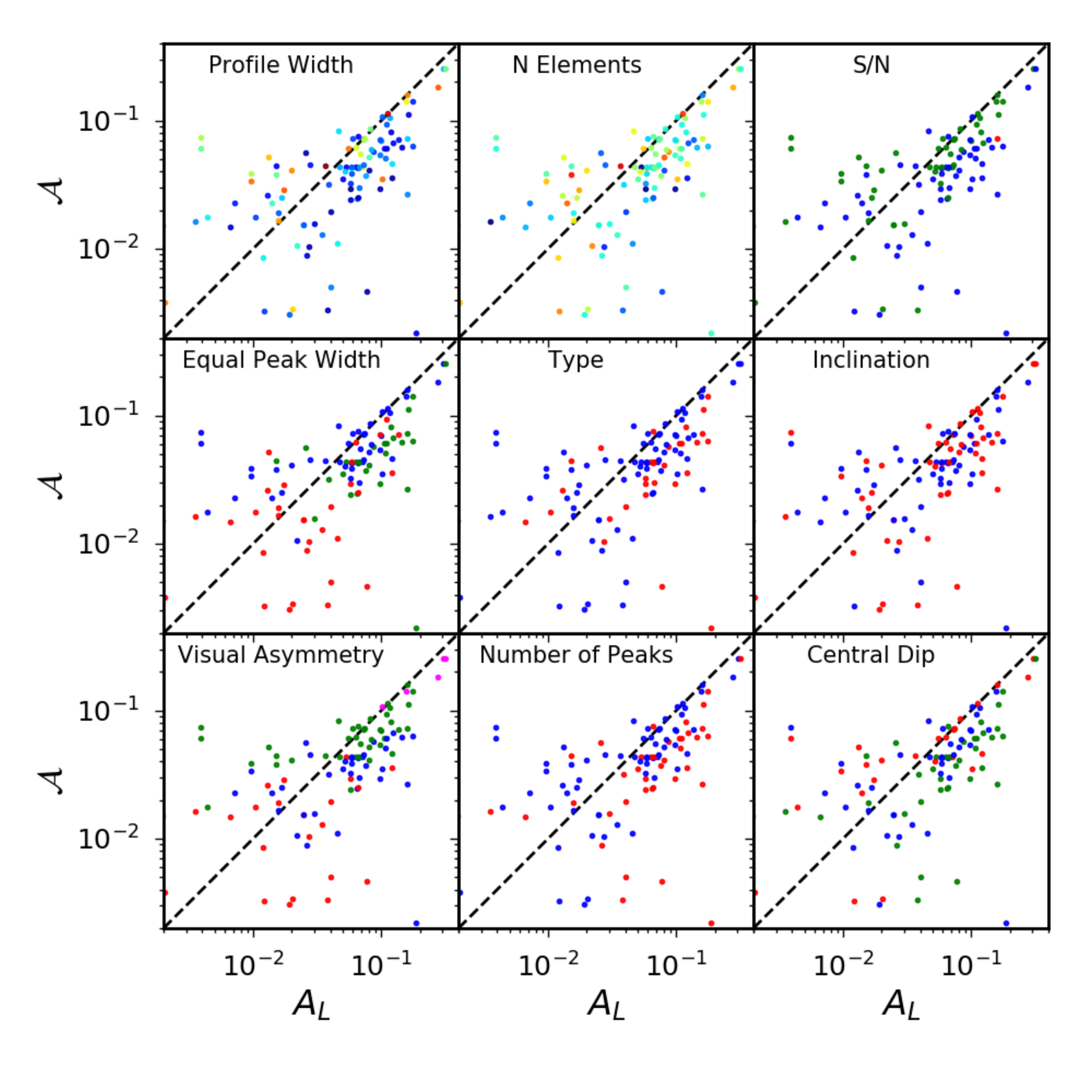}
\caption{Lopsidedness-Asymmetry correlations. Panels and point
colors are the same as Fig. \ref{Fig:LogDVLopCorr} 
\textcolor{black}{except that the S/N panel is now in 
the upper right.  The upper middle panel shows the 
points colored by the number of channel elements within
the profile where few to many channels goes from 
blue to red in the same manner as the profile width panel.}}
  \label{Fig:LogAsymLopCorr}
\end{figure*}

Figure \ref{Fig:LogAsymLopCorr} 
focuses on the  $\mathcal{A}$-$A_{L}$
correlation and its relation to 
other parameters.  This plot shows a number of features.
Firstly, profiles with a larger width tend to 
have larger values of $\mathcal{A}$.  This result points
to a profile resolution effect. 
\textcolor{black}{
However, this is not
completely clear as WHISP profiles have 
velocity resolutions of 
2, 4, 8, or 16 km s$^{1}$ depending on their global profile
width.  For this reason we've added the number of elements
in the upper middle panel of Fig. \ref{Fig:LogAsymLopCorr}.
That panel shows more clearly that profiles with 
more channels tend to have higher values of 
$\mathcal{A}$ for a given value of $A_{L}$.  This 
is consistent with Fig. \ref{Fig:ResolutionDependence},
which shows
that when the number of channels in a profile is 
below some limit (generally 20), $\mathcal{A}$ tends to 
decrease while the uncertainties increase.  The 
$\mathcal{A}$-profile width result seen 
in Fig. \ref{Fig:LogAsymLopCorr} is possibly a 
the manifestation of this trend.  It points to the 
need to be careful in both interpreting $\mathcal{A}$
distributions and comparisons between specific profiles.
}

The trend with profile width also explains the trends seen in 
the galaxy-type, number of peaks, and central dip panels of 
Fig. \ref{Fig:LogAsymLopCorr}.  The spiral galaxies tend 
to have larger widths than the irregulars.  It is also less
likely to get double-peaked profiles with narrow widths.  And, 
as discussed for the profile 
width trends seen in Fig. \ref{Fig:LogDVLopCorr}, narrow profiles
tend to have shallower central dips.

The \bsq{Equal Peak Width} panel in Fig. \ref{Fig:LogAsymLopCorr}
shows that double peaked profiles with different peak
widths tend to have the largest values of $\mathcal{A}$, while
they are spread across all values of $A_{L}$.  This result
is due to the fact that $A_{L}$ calculates the global
flux ratios while $\mathcal{A}$ is sensitive to how 
the flux is distributed on a channel-by-channel basis.  This same
trend is seen, but on a weaker level, for single peaks with 
differing slopes.  

\textcolor{black}{As with Fig. \ref{Fig:LogDVLopCorr}, there
is no clear trend with inclination.  There is a clear
trend with S/N, where galaxies that have higher
S/N values tend to have larger $\mathcal{A}$ for the 
same $A_{L}$.  This trend can be understood from the
results of Fig. \ref{Fig:NoiseResponse}.  There $A_{L}$ rapidly decreases to an asymptotic 
value, but, due to the background subtraction,
$\mathcal{A}$ has a slow increase with S/N.
As such, it is unsurprising that real galaxies with 
less noise tend to have larger values for $\mathcal{A}$.
It is perhaps best to think of $\mathcal{A}$ as a minimum
due to the amount of 
asymmetry in the signal region that may be attributed 
to the background.}

\textcolor{black}{Finally, this figure also shows the 
same result as Fig. \ref{Fig:HistogramComparison};
namely that $\mathcal{A}$ correlates better with the 
by-eye classification than $A_{L}$.}
Generally, the profiles classified
as moderately or strongly asymmetric have larger values of 
$\mathcal{A}$ than the slightly asymmetric or symmetric profiles.
Conversely, there are slightly asymmetric profiles with 
fairly large values for $A_{L}$ and 
moderately asymmetric profiles with quite low
$A_{L}$ values.  In other words, $\mathcal{A}$ better
represents what we perceive to be asymmetric.  However, it
is important to note that both $\mathcal{A}$ and our 
visual classification are biased towards wider profiles.
This is due to wider profiles having more channels and 
are therefore able to show more locally asymmetric features.

\section{Conclusions}\label{Sec:Conclusions}

In this work we have developed two new general statistics
to quantify the asymmetry of a velocity profile;
$\mathcal{A}$ and $\Delta V$.
Of these, $\mathcal{A}$ can be calculated at both 
the systemic velocity and the \bsq{velocity of symmetry}.
We have compared all three of these to the more 
commonly used lopsidedness statistic, $A_{L}$.

Using test profiles generated from the sum of 
Gaussian profiles, we explored how all the asymmetry 
statistics depend on $v_{fold}$ and $S/N$.  The lopsidedness
statistic is defined in such a way that it always has a 
minimum value of zero.  Conversely $\mathcal{A}$,  has a 
unique minimal value which depends on the profile.  This 
minimum is located at $v_{sym}$, which is not necessarily 
equal to $v_{sys}$.  From these results we conclude that 
$A_{L}$ should always be calculated at $v_{sys}$.  The 
channel-by-channel asymmetry \textcolor{black}{may} be calculated at
either $v_{sys}$ or $v_{sym}$.

Lopsidedness and $\Delta V$ decrease to some asymptotic value
as the $S/N$ increases.  The asymmetry statistic has the 
opposite behaviour due to the subtraction of a background
term.  All the statistics rely on both sufficient
$S/N$ and resolution elements.  In general, at low 
$S/N$, $\sim20$ channels is necessary for a reliable
asymmetry calculation.  However, at higher $S/N$ ratios,
accurate measurements can be made with even $10-15$ 
elements.

We also explored the dependence of the three statistics
on viewing angle and inclination using two snapshots from
simulations of interacting galaxies.  The fact that a 
velocity profile is a combination of the projection of 
kinematic and morphological asymmetries along a particular
line of sight means that profiles depend strongly on
the viewing angle.  As such, the four statistics
vary significantly as the viewing angle changes.  However, 
$\mathcal{A}$ is the most stable against viewing angle
variations.
In both snapshots, for certain viewing angles, $A_{L}=\Delta V=0$,
even through the underlying gas dispersion is objectively asymmetric. \textcolor{black}{Nonetheless, the viewing angle
result points to the difficulty in interpreting asymmetry
statistics for a single profile.  While an asymmetric profile
almost certainly indicates that the galaxy is disturbed, a
symmetric profile does not necessarily indicate an 
undisturbed galaxy.  So it is possible to make 
conclusions about \textit{individual} asymmetric 
galaxies, it is not possible to do so for symmetric 
galaxies.  But, it is possible to apply the statistics 
to large samples and make conclusions about \textit{populations} of symmetric and 
asymmetric galaxies..}

Similarly, $\mathcal{A}$ is also slightly more stable against
inclination.  While $A$ and $\Delta V$ vary significantly in 
the Shock Sim, $\mathcal{A}$ is constant for most inclinations.  It is 
worth noting that any dependence on inclination in this 
test is also degenerate with a dependence on 
resolution.  As the vertical motions are relatively 
small, the dominant effect of changing the inclination is 
to change the number of bins spanning the profile.  
In the Fly-by Sim, the vertical 
motions lead to significant effects on $\mathcal{A}$ and 
$\mathcal{A}(v_{sys})$ with peaks occurring 
as the profile becomes more face-on.  \textcolor{black}{Based
on our results we would suggest only applying these
statistics to galaxies with inclinations above
$15-20^{\circ}$.}

Finally, we applied all three statistics to a subsample
of real profiles drawn from WHISP.  We first visually 
classified each profile according to their number of 
peaks, apparent asymmetry, central dip, and the shape of 
the peaks.  We found that $A_{L}$ and $\Delta V$ are 
strongly correlated.  We also found that 
for a given value of $A_{L}$, profiles with deeper
central dips have larger values of $\Delta V$.  This means
that the $A_{L}-\Delta V$ space can be used to objectively select
profiles with or without deep central dips, which  
also points to the 
possibility of using a combination of asymmetry parameters
as a classification and selection tool for large 
upcoming surveys.  However, it will be necessary to determine
how the features identified by such a classification relate
to the actual galaxy.

We also found $\mathcal{A}$ is more 
strongly correlated with the visual classification than 
either of the other statistics.  In other words, 
the \textcolor{black}{channel-by-channel} asymmetry statistic
\textcolor{black}{calculated at the 
\bsq{velocity of symmetry} is} a superior indicator of 
non-axisymmetric features in velocity profiles than the 
lopsidedness or $\Delta V$ statistics.

The purpose of this paper was to introduce two new 
methods of characterizing velocity profiles.  These
methods, combined with the lopsidedness are 
promising new tools for upcoming surveys.  The
asymmetry statistics 
are sensitive to profile features not captured in the 
$v_{sys}$, profile width, and integrated flux.
Such features may arise from a variety of effects, 
like environment, merger histories, the method 
of gas accretion, etc.  

The variation of the different statistics on viewing
angle suggests that the measured values are not
necessarily reliable for a single galaxy.  However,
when they are applied to a sample they can 
give information about the populations.  This 
is partially seen in our work on the WHISP sample.

Our approach has been to develop profile 
characterization methods first and to understand how
they depend on a variety of unavoidable observational 
effects.  
\textcolor{black}{The intention behind these 
new asymmetry definitions is to use them to 
identify the physical state/processes of
a galaxy or population of galaxies (with
the caveat that symmetric profiles do not
always indicate undisturbed galaxies).  But, in order
to make conclusions about an object or population
of objects it is necessary to understand the contributions
from observational effects like S/N, resolution,
inclination and viewing angle. This is what 
we accomplished in Secs. \ref{Sec:SimpleProfiles} 
and \ref{Sec:Simulations}.  In the future we 
will explore the relationship between these profile
statistics and 2D morphological statistics.  We will 
also explore how they vary as a function of wavelength, and 
what correlations they have with the galaxy 
properties themselves (environment, star formation, etc.).
Nonetheless, it is 
clear from Sec. \ref{Sec:WHISP}
that profile asymmetry 
statistics do provide a useful method of characterizing
galaxies} and should be used to help analyze the
sea of data produced by upcoming deep HI surveys like LADUMA and DINGO.

\section*{Acknowledgements}

The authors wish to thank J. Bok for useful suggestions
and inspiration.  They would like to thank L. Cortese, 
A. Watts, L. Verdes-Montenegro , and K. Spekkens for their discussions.
They would also like to thank T. Jarret, A. Comrie, 
A. Sivitelli, and the IDIA data visualisation lab for 
allowing us to view the simulations in VR. 
\textcolor{black}{We also wish
to thank the referee for excellent suggestions
for improving this paper.}
ND's  work is supported by a SARChI South African 
SKA Fellowship.
NH  acknowledges  the  bursary  provided  by  South  African  Radio Astronomy Observatory.
SK  acknowledges  the  bursary  provided  by  the NRF via the National Astrophysics and Space Sciences Programme.
The numerical simulations were performed at the
Centre for High Performance Computing.

%%%%%%%%%%%%%%%%%%%%%%%%%%%%%%%%%%%%%%%%%%%%%%%%%%



%\bibliographystyle{mnras}

\begin{thebibliography}{99}

\bibitem[\protect\citeauthoryear{Angiras et al.}{2006}]{Angiras2006}
    Angiras, R.~A., Jog, C.~J., Omar, A., Dwarakanath, K.~S., 
    2006, MNRAS, 369, 	1849

\bibitem[\protect\citeauthoryear{Abazajian}{2009}]{Abazajian2009}
    Abazajian K. N., et al., 
    2009, ApJS, 182, 543

\bibitem[\protect\citeauthoryear{Arp}{1966}]{Arp1966}
    Arp, H.,
    1966, AJ Suppl. 14, 1.

\bibitem[\protect\citeauthoryear{Bloom et al.}{2017}]{Bloom2017} 
    Bloom J.~V., et al., 
    2017, MNRAS, 465, 123

\bibitem[\protect\citeauthoryear{Bloom et al.}{2018}]{Bloom2018} 
    Bloom J.~V., et al., 
    2018, MNRAS, 476, 2339
    
\bibitem[\protect\citeauthoryear{Blyth et al.}{2016}]{LADUMA} 
    Blyth S., et al., 2016, 
    mks..conf,  4, mks..conf 
    

\bibitem[\protect\citeauthoryear{Bok et al.}{2019}]{Bok2019} 
    Bok J., Blyth S.-L., Gilbank D.~G., Elson E.~C., 2
    019, MNRAS, 484, 582
    
\bibitem[\protect\citeauthoryear{Bournard et al.}{2005}]{Bournard2005} 
    Bournaud, F., Combes, F., Jog, C.~J., Puerari, I.,
    2005, A\&A, 438, 507    

\bibitem[\protect\citeauthoryear{Catinella et al.}{2018}]{Catinella2018}
	Catinella, B., et al.
	2018, MNRAS, 476, 875
	
\bibitem[\protect\citeauthoryear{Catinella et al.}{2010}]{Catinella2010}
	Catinella, B., et al.
	2010, MNRAS, 403, 638	


\bibitem[\protect\citeauthoryear{Conselice}{2003}]{Conselice2003}
	Conselice, C.~J.,
	2003, ApJS, 147, 1
	
\bibitem[\protect\citeauthoryear{Conselice et al.}{2000}]{Conselice2000}
	Conselice, C.~J., Bershady, M.~A., Jangren, A.,
	2000, ApJ, 529, 886	

\bibitem[\protect\citeauthoryear{Deg et al.}{2019}]{Deg2019}
	Deg, N., Widrow, L., Randriamampandry, T., Carignan, C.,
	2019, MNRAS, 486, 5391
	
\bibitem[\protect\citeauthoryear{Duffy et al.}{2012}]{WALLABYDINGO} 
    Duffy A.~R., et al.,
     2012, MNRAS, 426, 3385 
	
\bibitem[\protect\citeauthoryear{Espada et al.}{2011}]{Espada2011}
    Espada D., Verdes-Montenegro L., Huchtmeier W. K., Sulentic
J., Verley S., Leon S., Sabater J., 
    2011, A\&A, 532, A117	

\bibitem[\protect\citeauthoryear{Fern{\'a}ndez Lorenzo et al.}{2013}]{FernandezLorenzo2013}
    Fern{\'a}ndez Lorenzo, M., Sulentic, J., Verdes-Montenegro, L., Argudo-Fern{\'a}ndez, M., 
    2013, MNRAS, 434, 325	

\bibitem[\protect\citeauthoryear{Giese et al.}{2016}]{Giese2016}
    Giese N., van der Hulst T., Serra P., Oosterloo T., 
    2016, MNRAS, 461, 1656
    
 \bibitem[\protect\citeauthoryear{Gunn \& Gott}{1972}]{Gunn1972}
    Gunn, J.~E., Gott, J.~R.~III, 
    1972, ApJ, 176, 1   
    
\bibitem[\protect\citeauthoryear{Hank et al.}{2020}]{Hank2020}
	Hank, N., Blyth, S., Deg, N.
	2020, in prep     

\bibitem[\protect\citeauthoryear{Haynes et al.}{1998}]{Haynes1998}
	Haynes, M.~P., Hogg, D.~E., Maddalena, R.~J., Roberts, M.~S., van Zee, L.,
	1998, AJ, 115, 62 	
	
\bibitem[\protect\citeauthoryear{Haynes et al.}{2018}]{Haynes2018}      Haynes, M.~P., Giovanelli, R., Kent, B.~R., et al. 
    2018, ApJ, 861, 49	

\bibitem[\protect\citeauthoryear{Holwerda et al.}{2011}]{Holwerda2011}	
    Holwerda B.W., Pirzkal N., de BlokW. J. G., Bouchard A., Blyth
S.-L., van der Heyden K. J., 
    2011, MNRAS, 416, 2437
    
 \bibitem[\protect\citeauthoryear{Jarvis et al.}{2016}]{MIGHTEE} 
 Jarvis M., et al., 2016, 
 mks..conf,  6, mks..conf
   	

\bibitem[\protect\citeauthoryear{Kere{\v{s}} et al.}{2002}]{Keres2002}
    Kere{\v{s}}, D., Katz, N. Weinberg, D.~H., Dav{\'e}, R.,
    2002, MNRAS, 363, 2

\bibitem[\protect\citeauthoryear{Kornreich et al.}{2000}]{Kornreich2000}
    Kornreich D. A., Haynes M. P., Lovelace R. V. E., van Zee L.,
    2000, AJ, 120, 139

\bibitem[\protect\citeauthoryear{Lelli et al.}{2014}]{Lelli2014} 
    Lelli, F., Verheijen, M., Fraternali, F., 
    2014, MNRAS, 445, 1694


\bibitem[\protect\citeauthoryear{Lotz et al.}{2004}]{Lotz2004}
    Lotz J. M., Primack J., Madau P., 
    2004, AJ, 128, 163

\bibitem[\protect\citeauthoryear{Matthews et al.}{1998}]{Matthews1998}
    Matthews L. D., van Driel W., Gallagher III J. S., 
    1998, AJ, 116, 1169
        

\bibitem[\protect\citeauthoryear{Meyer et al.}{2004}]{Meyer2004}
    Meyer, M. J., Zwaan, M. A., Webster, R. L., Staveley-Smith, L., et al., 
    2004, MNRAS, 350, 1195 
    
\bibitem[\protect\citeauthoryear{Mundy et al.}{2017}]{Mundy2017}
    Mundy, C., J., Conselice, C., J., Duncan, K., J., Almaini, O. H{\"a}u{\ss}ler, B. Hartley, W., G., 
    2017, MNRAS, 470, 3507    
    
\bibitem[\protect\citeauthoryear{Newville et al.}{2014}]{LMFIT}
    Newville, M., Stensitzki, T., Allen, D.~B., Ingargiola, A.,
    2014, LMFIT: Non-Linear Least-Square Minimization and Curve-Fitting for Python, Zenodo, doi:10.5281/zenodo.11813

\bibitem[\protect\citeauthoryear{Noordermeer et al.}{2005}]{Noordermeer2005}
	Noordermeer, E., van der Hulst, J.~M., Sancisi, R., Swaters, R.~A., van Albada, T.~S.,
	2005, A\&A, 442, 137 


\bibitem[\protect\citeauthoryear{Peterson \& Shostak}{1974}]{Peterson1974}
    Peterson, S. D., Shostak, G. S.,
    1974, AJ, 79, 767
    
\bibitem[\protect\citeauthoryear{Reichard et al.}{2008}]{Reichard2008}
	Reichard, T.~A., Heckman, T.~M., Rudnick, G., Brinchmann, J., Kauffmann, G.,
	2008, ApJ, 677, 186	    

\bibitem[\protect\citeauthoryear{Richter \& Sancisi}{1994}]{Richter1994}
	Richter, O.-G., Sancisi, R.,
	1994, A\&A, 290, L9	
	
\bibitem[\protect\citeauthoryear{Sancisi, et al.}{2008}]{Sancisi2008} 
    Sancisi R., Fraternali F., Oosterloo T., van der Hulst T., 
    2008, A\&ARv, 15, 189	
    
\bibitem[\protect\citeauthoryear{Schade et al.}{1995}]{Schade1995}
	Schade, D., Lilly, S.~J., Crampton, D., Hammer, F., Le Fevre, O., Tresse, L.,
	1995, ApJ, 451, L1    
    
\bibitem[\protect\citeauthoryear{Shapiro et al.}{2008}]{Shapiro2008}           
    Shapiro K.~L., et al., 
    2008, ApJ, 682, 231

\bibitem[\protect\citeauthoryear{Springel}{2005}]{Springel2005}
	Springel, V.,
	2005, MNRAS, 464, 1105  
	
\bibitem[\protect\citeauthoryear{Stewart, Blyth \& de Blok}{2014}]{Stewart2014} 
    Stewart I.~M., Blyth S.-L., de Blok W.~J.~G., 
    2014, A\&A, 567, A61	
	
\bibitem[\protect\citeauthoryear{Swaters et al.}{2002}]{Swaters2002}
	Swaters, R.~A., van Albada, T.~S., van der Hulst, J.~M., Sancisi, R.,
	2002, A\&A, 390, 829 	

\bibitem[\protect\citeauthoryear{Tifft \& Cocke}{1988}]{Tifft1988} 
    Tifft W.~G., Cocke W.~J., 
    1988, ApJS, 67, 1
    

\bibitem[\protect\citeauthoryear{Trujillo et al.}{2001}]{Trujillo2001}
    Trujillo I., Aguerri J. A. L., Cepa J., Guti\'{e}rrez C. M., 
    2001, MNRAS, 328, 977


\bibitem[\protect\citeauthoryear{van der Hulst, van Albada \& Sancisi}{2001}]{vanderHulst2001} 
    van der Hulst J.~M., van Albada T.~S., Sancisi R., 
    2001, ASPC,  451, ASPC..240

\bibitem[\protect\citeauthoryear{van Eymeren et al. }{2011}]{vanEymeren2011} 
    van Eymeren, J., J{\"u}tte, E., Jog, C.~J., Stein, Y., Dettmar, R. -J.,
       2011, A\&A,  530, A29
       

\bibitem[\protect\citeauthoryear{Watts et al.}{2020}]{Watts2020} 
    Watts A.~B., Catinella B., Cortese L., Power C., 2020, MNRAS, 492, 3672

\bibitem[\protect\citeauthoryear{Westmeier, et al.}{2014}]{Westmeier2014}             Westmeier T., Jurek R., Obreschkow D., Koribalski B.~S., Staveley-Smith L.,       2014, MNRAS, 438, 1176


\bibitem[\protect\citeauthoryear{Wisnioski et al.}{2019}]{Wisnioski2019}
    Wisnioski, E., et al.
    2019, ApJ, 886, 124

\bibitem[\protect\citeauthoryear{Wilcots \& Prescott}{2004}]{Wilcots2004}
    Wilcots E.~M., Prescott M.~K.~M., 
    2004, AJ, 127, 1900



\end{thebibliography}


% Don't change these lines
\bsp	% typesetting comment
\label{lastpage}
\end{document}